\documentclass{amsart}

\usepackage{graphicx}        

\usepackage{amsmath,amsfonts}%
\usepackage{amsthm}%
\usepackage{xcolor}%
\usepackage{algorithm}%
\usepackage{algorithmicx}%
\usepackage{algpseudocode}%
\usepackage{comment}
\usepackage{subcaption}
\usepackage{psfrag}
\usepackage{enumerate}
\usepackage{tikz}
\usepackage[margin=1in]{geometry}
\usepackage{url}

\newtheorem{definition}{Definition}
\newtheorem{remark}{Remark}
\newtheorem{theorem}{Theorem}
\newtheorem{lemma}{Lemma}
\newtheorem{proposition}{Proposition}
\newtheorem{observation}{Observation}

\newtheorem*{conjecture*}{Conjecture}

\title{Cayley Configuration Spaces of a Common Class of Mechanisms in Two Dimensions}
\author{Meera Sitharam$^1$}
\author{Menghan Wang$^2$}
\author{William Sims$^{3,*}$}
\author{Heping Gao$^4$}
\thanks{$^1$CISE Department, University of Florida, Gainesville, FL, USA}
\thanks{$^2$Netflix}
\thanks{$^3$School of Information Technology, Illinois State University, Normal, IL, USA}
\thanks{$^*$Corresponding author, wsims3@ilstu.edu}
\thanks{$^4$Meta}
\date{}

\begin{document}

\begin{abstract}
 
We study Cayley configuration spaces of a class of 1 degree-of-freedom linkages (graphs with specified edge lengths), obtained by dropping an edge from a tree-decomposable graph. The class includes well-known  mechanisms based on the four-bar, as well as strandbeest, cardioid, limacon etc. 
The Cayley configuration space is the set of intervals of attainable
 lengths for a \emph{base} nonedge  (e.g. the dropped edge) over the linkage's 2-dimensional  
realizations.  We require \emph{quadratic radical solvability (QRS)} (an extension of ruler-and-compass-realizability) of   the interval endpoints, and tree-decomposability guarantees efficient, ruler-and-compass construction of the linkage realization, given the Cayley configuration.  
 Due to these restrictions of Kempe universality, this class of \emph{low Cayley complexity (LCC)} graphs  is common in mechanical computer aided design and kinematics. Our main contributions are the following.
 (1) We show that the definition of LCC is robust, and depends only on the graph, no matter  
 the choice of base nonedge whose addition ensures tree-decomposability. 
 (2) We give an efficient algorithmic characterization of LCC graphs 
 (3)  We show (graph) planarity is equivalent to
 LCC for a natural subclass of 1-degree-of-freedom 
tree-decomposable graphs.
Counterexamples show impossibility of such finite forbidden minor characterizations
when the above subclass is enlarged.
(4)  We give an easily testable definition of genericity of LCC linkages (i.e. with underlying LCC graphs)  based on their edge lengths. 
 (5) For generic LCC linkages,  we give an algorithm to
find both paths of continuous motion
(provided they exist) between two  distinct realizations, in time linear in a discrete measure
of the length of the path.
 Nontrivial  generalizations of these results  to non-LCC, 1-degree-of-freedom tree-decomposable linkages. Several accessible open problems are posed.
 \end{abstract}

\maketitle

\section{Introduction}
\label{sec:introduction}
A {\em linkage} $(G,\bar{l})$, is a graph $G=(V,E)$ with fixed
length bars as edges, i.e. $\bar{l}: E \rightarrow \mathbb{R}$. 
A (2-dimensional Cartesian) 
\emph{realization} $G(p)$ of the linkage $(G.\bar{l})$ is an assignment of points $p \in \mathbb{R}^2$ to the vertices of $G$, 
satisfying the bar lengths in $\bar{l}$, i.e., for all edges $(u,v) \in G$, $\| p_u - p_v \| = \bar{l}(u,v)$ where $p_u,p_v \in \mathbb{R}^2$. 
Note that a linkage may or may not have a 2D realization. 

We will use standard and well-known terminology from geometric constraint solving and combinatorial rigidity. 
For detailed background in these fields we refer the reader to, for example,
\cite{bib:Graver,sitharam2018handbook,bib:survey}.
In general, properties such as rigidity, flexibility and independence  of a linkage refer to the corresponding properties of the underlying graph in 2D, ignoring the edge lengths. This will be justified by an important contribution of this paper: a testable definition of genericity of a linkage based on the edge lengths.  
 The \emph{degrees of freedom (dofs)} of a linkage
 is the minimum number of edges that must be added to the graph of the linkage to make it rigid.

 We restrict ourselves to 1-dof linkages  (called \emph{mechanisms} in the kinematics literature) 
obtained by dropping a bar from minimally rigid, \emph{tree-decomposable linkages} (studied in this paper and defined formally in Section \ref{sub:definitions}). This class of mechanisms is 
 widely used in engineering and CAD, 
because they are ruler-and-compass realizable or \emph{quadratically-radically solvable} (\emph{QRS}),
i.e., given the edge lengths in $\mathbb{Q}$, the realizations are solutions to a
triangularized quadratic system with coefficients in $\mathbb{Q},$
i.e. the  vertex coordinate values in the realization belong to an extension field over
$\mathbb{Q}$ obtained by nested square-roots, we refer to the elements of this field as QRS as well.  
Conversely, QRS has been shown \cite{bib:Owen02} to generically imply tree-decomposability
in the case of  planar graphs, and the implication is strongly conjectured for
all graphs. 

The paper studies realization spaces of a common class of 1-dof tree-decomposable linkages.

\subsection{State of the Art and Motivation}

Describing and analyzing realization spaces of 1-dof linkages
 is 
a  difficult problem even in 2D, with a long history. 
In mechanical computer aided design, it represents a key underlying barrier
to understanding underconstrained geometric constraint systems.
In fact, even for rigid linkages, the number of realizations can be exponential in the number of vertices
and not easy to estimate \cite{bib:Borcea,capco2018number,gallet2019counting}. 
One way to classify realizations is to use \emph{realization types} \cite{bib:navigation,bib:FudHo97}, used extensively in this paper and formally defined  in Section \ref{sub:definitions}.
which can uniquely determine a realization of a rigid linkage. 
For flexible linkages, 
a well-known early result \cite{kempe1875} 
shows that an arbitrary algebraic curve can be traced by the motion of a linkage joint. 
One outstanding example is the Peaucellier-Lipkin linkage, which transforms planar rotary motion into straight-line motion \cite{bib:Kempe}. Other well-known examples include the four-bar, strandbeest, limacon and carioid linkages \cite{bib:beast,bib:caymos}.
For polygonal linkages, recent results on the variants of Carpenter's rule problem and pseudotriangulations yield spaces of non-crossing realizations and expansive motions \cite{bib:streinu05,streinu2000combinatorail,bib:straightening,bib:rote03}.
Versions of the problem play an important role in Computer-Aided-Design (CAD), 
robotics and molecular geometry \cite{bib:sacks10,bib:Yang,bib:survey}, but few results are known beyond individual or specific families of linkages \cite{bib:JoanArinyo03,bib:hilderick06,bib:henneberg,bib:ZhangGao06}. 

There are numerous examples of algorithms and software dealing with 1-dof linkages, 
such as Geometry Expressions, SAM,  Phun, Sketchpad, Geogebra, D-cubed, 
the algorithm in \cite{bib:hidalgo2011reachability}, etc. 
They have the following major functionalities: 
(i) designing 1-dof  linkage  for tracing out specific curves, 
especially by building new  linkages based on a library of existing ones;
(ii) accepting user-specified parameters, ranges and realization types to generate 
continuous motion of the linkages.

However, the following critical issues  hinder the state of the art. 

(a) There is an  essential and judicious \emph{interplay
between simple algebraic geometry properties}  
\emph{and simple graph theoretic properties} 
of  1-dof linkages.
However, to the best of our knowledge, the only known formal result in this area that has a similar flavor
of combinatorially capturing algebraic complexity is the result of \cite{bib:Owen02} that relates quadratic
solvability and tree-decomposability for planar graphs.

(b)Concerning  parametrizations or other efficient descriptions of  the  realization space
space, the papers \cite{bib:JoanArinyo03,bib:survey,bib:ZhangGao06} study how to obtain ``completions'' 
of  flexible
graphs $G$, i.e, a set of nonedges $F$ whose addition makes $G$ minimally rigid. 
All are motivated by the need to efficiently obtain realizations of flexible linkages. 
In particular \cite{bib:JoanArinyo03} also guarantees that the completion ensures tree-decomposability.
However, they do not even attempt to address the question of how to find realizable
distance values for the completion edges. Nor do they address
algebraic complexity of the set of distance values that these completion nonedges
can take, nor the complexity of obtaining a description of this configuration space,
nor a combinatorial characterization of graphs
for which this complexity is low. 
The latter factors however are crucial for tractably analyzing and decomposing 
the realization space in order to obtain the corresponding realizations. 
 A single exception is  
the paper \cite{bib:hilderick06} gives a collection of useful observations and heuristics for computing
 extreme points in the  realization space descriptions of certain linkages that
arise in real CAD applications, by decomposing the linkages into subproblems.
However, it relies on a complete list of solutions for all possible subproblems.
Nor does it address the complexity issues mentioned above.

(c) Currently, 
the realization space is typically represented as separate curves in 2D that are 
traced by each vertex of the linkage. 
In fact, 
a realization actually corresponds to a tuple of points, one on each of these curves.
I.e., the realization space  maps bijectively to a curve in the full 
\emph{ambient dimension} of $2|V|-3$ after factoring out rigid transformations, 
where $|V|$ is the number of vertices in the linkage. 

(d) Currently, 
for two realizations in different connected components, 
there is no method to find out how ``close" they can get towards each other by continuous motion, 
using a meaningful definition of ``distance" between connected components. 

(e) Currently, in order to generate continuous motion, 
the user must 
specify a range of a parameter containing the parameter value at the given realization.
Then either a single connected component is generated
for a subset of the specified range, 
or multiple segments of the realization space, under only the given realization type,  
are generated within the specified range. 
We discuss this issue in more detail in Section \ref{sub:relation}.


\subsection{Organization}
\label{sec:organization}
In  Section \ref{sec:contributions}, we give basic definitions, state our main contributions and  their relation to previous results.  

In Section \ref{sec:Basic-properties}, we give basic definitions 
related to 1-dof tree-decomposable graphs. 
In Section \ref{sec:Combinatorial-interpretation-of}, we associate a special so-called \emph{extreme graph} with each interval endpoint of a (oriented) Cayley configuration space and give the precise definition of the LCC class. 

In Section \ref{sec:Characterizing-parameter-choices:}, we prove that the definition of LCC is robust as a graph property, i.e. 
the choice of base nonedge does not affect LCC. 

In Section \ref{sub:Characterizing-general}, we give a characterization
of  LCC, which we call the \emph{four-cycle} characterization,
yielding an efficient recognition algorithm.   

In Section \ref{sub:Forbidden-minor-characterization}, 
we study finite forbidden minor characterizations of low Cayley (algebraic) complexity. 
We introduce two natural subclasses: 1-path and
$K_3$-free. 
In Section \ref{sub:Forbidden-minor-characterization}, we prove that LCC is equivalent to planarity for 1-path, $K_3$-free, 1-dof
tree-decomposable graphs. 
In Section \ref{sec:tightness}, we show that there is no finite forbidden minor characterization for more general classes of graphs of low Cayley (algebraic) complexity. 

In Section \ref{sec:Find-Cayley-configuration}, we  describe
  a natural \emph{minimal realization type} -- i.e. a minimal set of local orientations --  
for an LCC linkage (i.e. a linkage with an LCC underlying graph), whose specification ensures that
 there is at most a single interval in the Cayley configuration space, and  a  quadratic time  algorithm to find its endpoints.
We also show minimality, i.e., that specifying fewer orientations fails to restrict the Cayley configuration space  to polynomially many intervals.   In fact  if the set of relative orientations is small enough, even deciding whether this oriented Cayley configuration space is non-empty does not have a polynomial time algorithm unless P = NP \cite{bib:saxe79}. In other words, listing all the interval (endpoints) could take exponential time even if the time complexity is measured w.r.t. output size, i.e. the number of nonempty intervals.

In Section \ref{sec:cont-path}, 
we give an algorithm to find a path of continuous motion between two given realizations   
and show that there are  two such paths.  
We also show that when those two realizations have the same minimal realization type, 
a path staying within the same minimal realization type can always be found in constant time. 

In Section \ref{sec:ambient}, we give a canonical bijective representation the the realization space, 
which yields a  measure of distance between connected components of the realization space, and a meaningful visualization of the realization space as curves in an ambient space of minimal dimension.

\section{Basic Definitions, Contributions, Previous Work}
\label{sec:contributions}
We first provide basic definitions in Section \ref{sub:definitions}  necessary for stating the  the main contributions in Section \ref{sub:contributions}, and their relation to previous work in \ref{sub:relation}.

\subsection{Basic Definitions}
\label{sub:definitions}

Here we define Cayley configuration spaces, realization types, quadratic radical solvability (QRS), 1-dof tree-decomposable graphs, their base nonedges, and low cayley complexity (LCC)  (informally) and genericity of 1-dof tree-decomposable linkages.   

 \subsubsection{Cayley Configuration Spaces of Linkages and Realization Types}
We study the \emph{Cayley configuration space} of a linkage, first introduced in \cite{bib:GaoSi05,bib:GaoSitharam08a} and formally defined  below,
which is a set of intervals of possible distance-values for an independent nonedge of the linkage.
Let $G  = (V,E)$.  
We write the graph $(V,E\cup\{f\})$, where the endpoints of $f$ are in $V$, as $G\cup f$.
A reasonable way to describe the space of 2D realizations of a 1-dof linkage $(G,\bar{l})$ is to take a pair of vertices whose distance is not fixed $\bar{l}$,
i.e., an {\em independent nonedge} $f$ with $G\cup f$ being minimally rigid, 
and ask for all the possible lengths $l_f$ that the nonedge $f$ can attain 
(i) over all the realizations of $(G, \bar{l})$; 
(ii) over all realizations of $(G, \bar{l})$ that agree with a particular \emph{$T$-realization type}. 
A \emph{$T$-realization type} of a 2D realization $G(p)$,
where $T$ consists of triples of points in $p$, 
is a set $\sigma$  
of \emph{local orientations} (chirality) $\sigma_t \in \{+1,-1,0\}$, each denoting the \emph{local orientation} of a specific 
triple of points $t = (p_u,p_v,p_w) \in T$ , i.e. the sign of the determinant $\left|\begin{array}{c}p_w-p_u\\p_v-p_u\end{array}\right|$.  
A realization agrees with $\sigma$ if each nonzero coordinate of its $T$-realization type matches the corresponding coordinate in $\sigma$.  

For (i) (resp. (ii)), we call each realizable length of $f$ as a (resp. \emph{$\sigma$-oriented}) \emph{Cayley configuration}, 
and the set of all such configurations as the (resp. \emph{$\sigma$-oriented}) \emph{Cayley configuration space} of the linkage $(G,\bar{l})$ over $f$.  
Either Cayley configuration space is  a set of disjoint closed intervals on the real line.
Aside: the Cayley configuration space over $f$ 
is actually the \textsl{projection} of the Cayley-Menger semi-algebraic set \cite{bib:cayley}
associated with the linkage $(G,\bar{l})$ on the Cayley nonedge length parameter $f$.

\begin{figure}[h]
	\psfrag{1}{$v_{1}$}\psfrag{2}{$v_{3}$}\psfrag{3}{$v_{2}$}\psfrag{4}{$v_{4}$}\psfrag{5}{$v_{5}$}\psfrag{6}{$v_{6}$}\psfrag{7}{$v_{7}$}\psfrag{8}{$v_{8}$}
	
	\begin{centering}
	\includegraphics[width=0.5\textwidth]{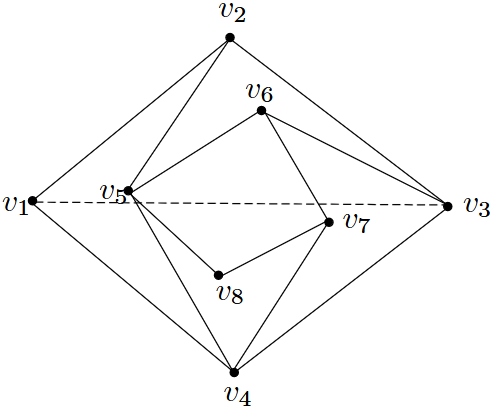}
	\par\end{centering}
	
	\caption{A 1-dof graph: adding any of the edges $(v_i,v_{1+2})$ would make the graph tree-decomposable. 
	In Figure \ref{F:interval}, we show the Cayley configuration space of a linkage with this graph over the dashed base nonedge $(v_1,v_3)$.  
    See the discussion below.  }

\label{F:intro} 
\end{figure}

\begin{figure}[h]
	
	\begin{centering}
	\includegraphics[width=\textwidth]{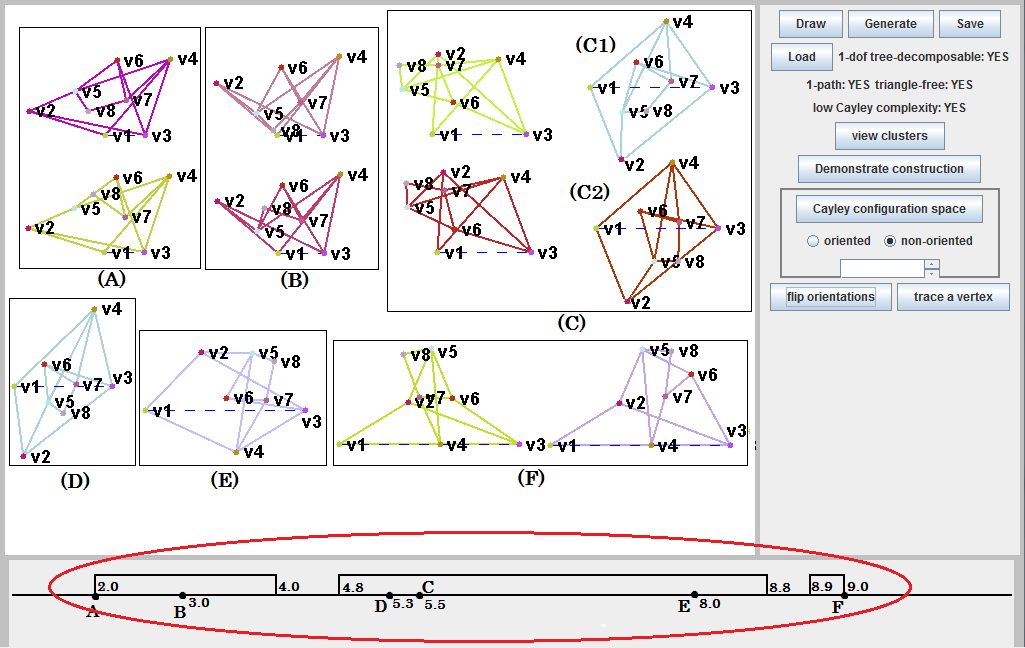}
	\par\end{centering}
	
	\caption{
	The circled part is the Cayley configuration space of a 1-dof tree-decompasable linkage, whose graph is shown in Figure \ref{F:intro}, over the base nonedge $f = (v_1,v_3)$ 
	(demonstrated using our  CayMos software \protect\cite{bib:caymos}).  
	Each point in this space corresponds to many realizations. 
	Arbitrarily chosen realizations for the various lengths $l_f$ for $f$: (A) $l_f=2$, (B) $l_f=3$, (C) $l_f=5.5$, (D) $l_f = 5.3$, (E) $l_f = 8$, (F) $l_f = 9$.  
    See the discussion below and Sections \ref{sub:relation} and \ref{sec:Combinatorial-interpretation-of}.}

\label{F:interval} 
\end{figure}

\begin{figure}[h]
	\psfrag{1}{$v_{1}$}\psfrag{2}{$v_{2}$}\psfrag{3}{$v_{3}$}\psfrag{4}{$v_{4}$}\psfrag{5}{$v_{5}$}\psfrag{6}{$v_{6}$}\psfrag{7}{$v_{7}$}\psfrag{8}{$v_{8}$}\psfrag{9}{$v_{9}$} 
	\psfrag{s1}{$\sigma$}\psfrag{s2}{$\tau$}
	
	\begin{centering}
	\includegraphics[width=0.7\textwidth]{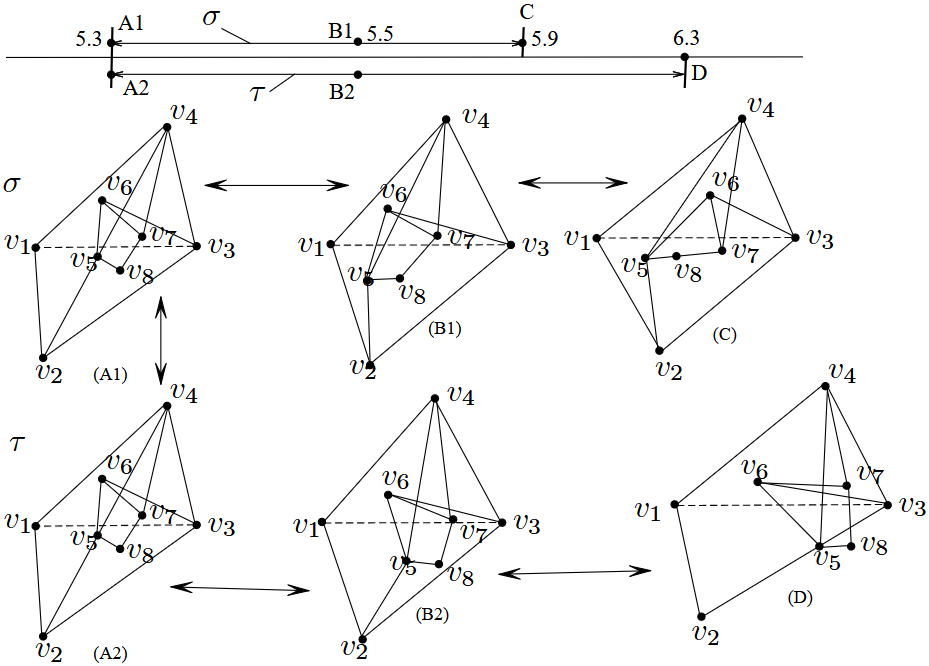}
	\par\end{centering}
	
	\caption{Two oriented Cayley configuration spaces of a 1-dof linkage, whose graph is shown in Figure \ref{F:intro}, over the base nonedge $f = (v_1,v_3)$, and   
	unique realizations for various lengths $l_f$ of $f$: (A1)(A2) $l_f = 5.3$, (B1)(B2) $l_f = 5.5$ (realizations (C1)(C2) from Figure \ref{F:interval}), (C) $l_f = 5.9$, (D) $l_f = 6.3$.  
    See the discussion below and Section \ref{sec:cont-path}.}

\label{F:tracking} 
\end{figure}

For example, the linkage $(G,\bar{l})$ in Figure \ref{F:intro} is 1-dof, 
and adding a bar between any of the pairs $(v_i, v_{i+2})$ would make it rigid.
In Figure \ref{F:interval},  
we choose $(v_1,v_3)$ as the nonedge $f$ of $(G,\bar{l})$ and
give the Cayley configuration space of this linkage over $f$,
consisting of the three intervals shown. 
It is important to note that Cayley configuration is not always a bijective representation, 
since each Cayley configuration can (generically) correspond to (finitely) many realizations: 
the figure shows arbitrarily chosen realizations corresponding to six Cayley configurations.
In Figure \ref{F:tracking}, 
we take \emph{forward realization types} (a special T-realization type defined in Definition \ref{def:solution-type})
$\sigma$ of realization (C1) from Figure \ref{F:interval} and $\tau$ of realization (C2) from Figure \ref{F:interval}, 
and show the two corresponding \emph{forward oriented Cayley configuration spaces} of $(G,\bar{l})$
over $f$.  
Each of these oriented Cayley configuration spaces consists of a single interval. 

 There are two desirable requirements 
on the Cayley configuration space. 
First, for each Cayley configuration $l_f$,
there should exist only finitely many (could be exponential in $|G|$) 
realizations of $(G\cup f,\bar{l})$. 
This is guaranteed if the linkage $(G\cup f, \bar{l})$ is rigid. 
Second, with a specified realization type, 
there should exist a linear time algorithm to convert 
from a Cayley configuration $l_f$ to a corresponding Cartesian realization, 
As an example, the linkage in Figure \ref{F:intro} satisfies both requirements 
when we choose any nonedge $f=(v_i, v_{i+2})$ as the Cayley parameter, 
since there exists a simple ruler and compass realization of the linkage $(G \cup f, \bar{l})$ from any such $f$. 
For linkages with such a realization, the coordinate values of realizations are QRS, 
and we call such linkages \emph{QRS linkages}, and the underlying graphs \emph{QRS graphs}.

With these two requirements in mind, 
we focus on a natural class of 1-dof linkages called \emph{1-dof tree-decomposable linkages}. 
The underlying graphs are obtained by dropping an edge from so-called \emph{tree-decomposable graphs} (formally defined  next).

\subsubsection{Tree-decomposable graphs}

Here we define concepts related to tree-decomposable graphs.  

\begin{figure}[h]
	\psfrag{A}{$v_{1}$} \psfrag{B}{$v_{2}$} \psfrag{C}{$v_{3}$}\psfrag{D}{$v_{4}$}
	\psfrag{E}{$v_{5}$} 
	\psfrag{1}{$G_{1}$} \psfrag{2}{$G_{2}$} \psfrag{3}{$G_{3}$}
	\psfrag{11}{$G_{11}$} \psfrag{12}{$G_{12}$} \psfrag{13}{$G_{13}$} 
	\psfrag{(a)}{(a)} \psfrag{(b)}{(b)}
	
	\begin{centering}
	\includegraphics[width=0.6\textwidth]{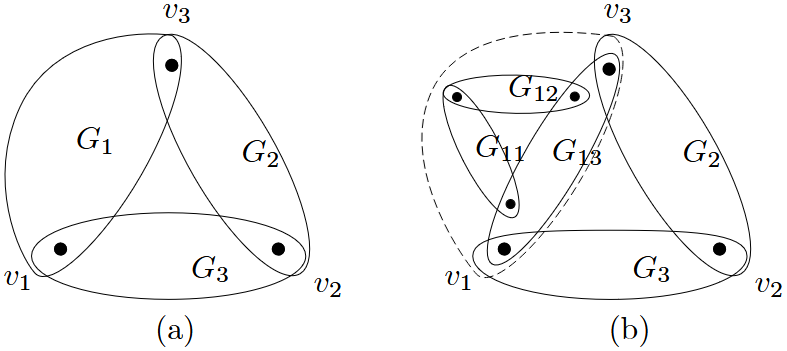} 
	\par\end{centering}
	
	\caption{A graph is tree-decomposable if it can be decomposed into three tree-decomposable subgraphs.  
    See the discussion below.}

\label{F:triangleDecomposition} 
\end{figure}

\begin{definition}[Tree-decomposable and 1-dof tree-decomposable graphs]
    \label{def:t-decomp}
    A graph $G$ is {\emph{tree-decomposable}} if either it is a single edge or it is the union of three tree-decomposable graphs such that each pair shares exactly one vertex and all three shared vertices are distinct (see Figure~\ref{F:triangleDecomposition}(a)).  
    $G$ is \emph{1-dof} tree-decomposable if it has some nonedge $f$ such that $G \cup f$ is tree-decomposable.  
    We call $f$ a \emph{base nonedge} of $G$ and a \emph{base edge} of $G \cup f$.  
\end{definition}

\begin{remark}
    Using using Laman's theorem \cite{bib:Laman70}, it is easy to see that a tree-decomposable graph is minimally rigid.  
    A 1-dof tree-decomposable graph $G$ can have many base nonedges. 
    That is, $G$ may have nonedges $f'\ne f$ such that both $G \cup f$ and $G\cup f'$ are tree-decomposable graphs. 
    We emphasize that this is different from deleting a different
    edge $e \neq f$ from $G\cup f$, which gives an entirely different 1-dof tree-decomposable graph from $G$.  
\end{remark}

For example, in Figure \ref{F:triangleDecomposition}, a tree-decomposable graph is decomposed into three tree-decomposable components, and $G_{1}$ is decomposed into $G_{11}$, $G_{12}$ and $G_{13}$.
Our initial example in Figure \ref{F:intro} is a 1-dof tree-decomposable linkage (i.e. a linkage whose graph is 1-dof tree-decomposable).
The following construction of 1-dof tree-decomposable graphs follows from the above definition.

\begin{figure}[h]
	\psfrag{1}{$v_{1}$} \psfrag{2}{$v_{2}$} \psfrag{3}{$v_{3}$}
	\psfrag{4}{$v_{4}$} \psfrag{5}{$v_{5}$} \psfrag{6}{$v_{6}$}
	\psfrag{7}{$v_{7}$} \psfrag{8}{$v_{8}$} \psfrag{0}{$v_{0}$}
	\psfrag{0'}{$v_{0}'$} \psfrag{t1}{$T_{1}$} \psfrag{t2}{$T_{2}$}
	\psfrag{t3}{$T_{3}$} \psfrag{t4}{$T_{4}$} \psfrag{t5}{$T_{5}$}
	\psfrag{t6}{$T_{6}$} \psfrag{t7}{$T_{7}$} \psfrag{t8}{$T_{8}$}
	\psfrag{(a)}{(a)} \psfrag{(b)}{(b)}
	
	\begin{centering}
	\includegraphics[width=\textwidth]{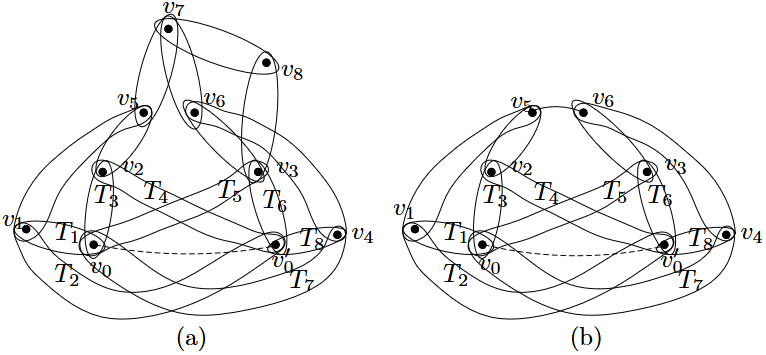} 
	\par\end{centering}
	
	\caption{(a) A 1-dof tree-decomposable graph with base nonedge $(v_{0},v_{0}')$ and levels $L_{0}$ to $L_{4}$. 
	(b) Extreme graph $\hat{G}_{f}(7)$ (defined in Section \ref{sec:Combinatorial-interpretation-of}) for (a).  
    See the discussion below and Section \ref{sec:Basic-properties}.}

    \label{F:construction} 
\end{figure}

\smallskip
\noindent\textbf{Construction.}
Any 1-dof tree-decomposable graph $G$ can be constructed iteratively as follows, starting from a given base nonedge $f$. 
At the $k^{th}$ {\emph{construction step}}, two new maximal tree-decomposable subgraphs
$T_{1}$ and $T_{2}$, called {\emph{clusters}}, sharing a single {\emph{step vertex}} $v_{k}$
are appended to the previously constructed graph $G_f(k-1)$: $T_{1}$
and $T_{2}$ each has exactly one shared vertex, $u_k$ and $w_k$ respectively, with $G_f(k-1)$, 
where $u_k \ne w_k$
(see Figure \ref{F:construction}). 
We denote this construction
step by $v_{k}\triangleleft(u_k\in T_{1},w_k\in T_{2})$, or simply $v_{k}\triangleleft(u_k,w_k)$.
 The  construction steps  can be arbitrarily ordered as long as the ordering is consistent with an underlying partial order on the step vertices, whose  the \emph{levels} are defined as follows: starting from   a base nonedge $f=(v_{0},v_{0}')$, 
each level is a set of vertices, where
\begin{itemize}
	\item Level $0$, denoted $L_{0}$  consists of $v_0$ and $v_0'$. 
	
	\item Level $1$, denoted $L_{1}$  is the set of vertices $v_i$ such that $v_i \triangleleft (v_0,v_0')$.
	
	\item Level $i$, denoted $L_{i}$  $(i \ge 2)$ is the set of vertices $v_i$ such that $v_i \triangleleft (u,v)$, 
		where $u \in T_j$, $w \in T_k$, with $j$ and $k$ being the minimum indices such that clusters $T_j$ and $T_k$  that were added along with a step vertex at level $j$ and $k$ respectively, and  $i-1$ is the maximum of $j$ and $k$.
\end{itemize}
A tree-decomposable graph can be constructed in a similar way from a given base edge.
From this construction, we get the following natural $T$-realization type.

\begin{definition}[Forward realization types]
    \label{def:solution-type}
    A \emph{forward realization type} of a 1-dof tree-decomposable graph $G$ is a $T$-realization type where, for some base nonedge $f$, $T$ consists of all triples $(u,w,v)$ where $v \triangleleft (u,w)$ is a construction step in the construction of $G$ from $f$.  
    A forward realization type is \emph{strict} if all its coordinates are nonzero.  
\end{definition}

\begin{figure}[h]
	\psfrag{T1}{$T_{1}$} \psfrag{T2}{$T_{2}$}\psfrag{T3}{$T_{3}$}\psfrag{T4}{$T_{4}$}
	
	\psfrag{T5}{$T_{5}$} \psfrag{T6}{$T_{6}$}\psfrag{T7}{$T_{7}$}\psfrag{T8}{$T_{8}$}
	
	\psfrag{v1}{$v_{0}$}\psfrag{v2}{$v_{0}'$}\psfrag{v3}{$v_{1}$}\psfrag{v4}{$v_{2}$}
	
	\psfrag{v5}{$w_{1}$}\psfrag{v6}{$v_{3}$}\psfrag{v7}{$w_{2}$}\psfrag{v8}{$v_{4}$}
	
	\psfrag{f}{$f$} 
	
	\begin{centering}
	\includegraphics[width=0.8\linewidth]{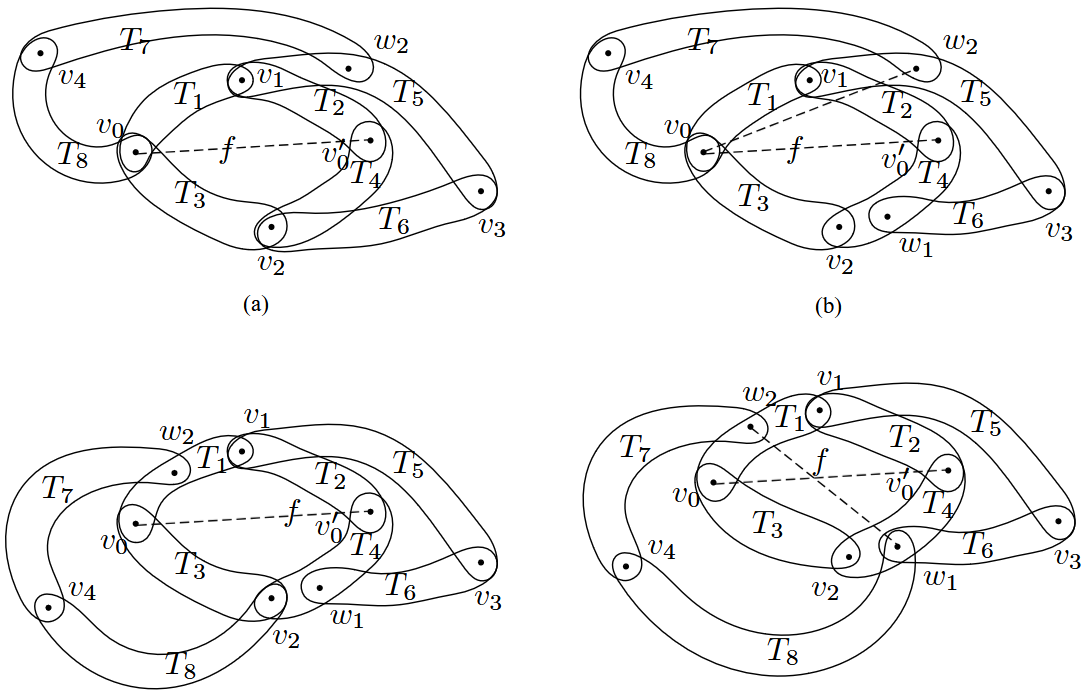} 
	\par\end{centering}
	
	\caption{(a)(b) are 1-path and (c)(d) are 1-dof tree-decomposable graphs that are not 1-path. (a)(c) are LCC while (b)(d) are not. 
    See the discussion below and Section \ref{sec:Combinatorial-interpretation-of}.}

\label{F:1-path} 
\end{figure}

\begin{figure}[h]
\psfrag{1}{$v_{1}$} \psfrag{2}{$v_{2}$}\psfrag{3}{$v_{3}$}\psfrag{4}{$v_{4}$} 

\psfrag{5}{$v_{5}$} \psfrag{f1}{$f_{1}$}\psfrag{f2}{$f_{2}$}

\begin{centering}
\includegraphics[width=.23\linewidth]{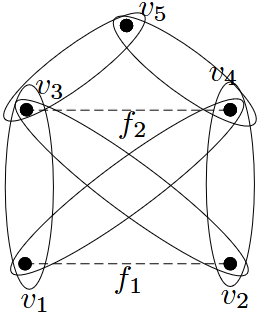} 
\par\end{centering}

\caption{Some choices of base nonedges have the same unique final level step vertex.  
See the discussion below.}

\label{F:1-path-base} 
\end{figure}

We give special attention to 1-dof tree-decomposable graphs that have constructions with a unique final step vertex.  

\begin{definition}[1-Path]
    \label{def:1-path}
    A \emph{1-path} graph $G$ is a 1-dof tree-decomposable with a \emph{1-path construction} from some base nonedge $f$: i.e., $G$'s construction from $f$ has a single step vertex and two clusters at the final level.  
\end{definition}

For example, consider the constructions of the graphs in Figure \ref{F:1-path} from the base nonedges $f=(v_{0},v_{0}')$.  
The constructions in (a) and (b) are 1-path since $v_{4}$ is
the only step vertex in the final level. 
The constructions in (c) and (d) are not 1-path since both $v_{3}$ and $v_{4}$ are step vertices in the final level.  
Note that a 1-dof tree-decomposable graph could have 1-path
construction from some base nonedges but not from others.  
For example, in Figure \ref{F:1-path-base}, the construction from $f_1$ is 1-path since $v_5$ is the only step vertex in the final level.  
On the other hand, the construction from $f_2$ is not 1-path since $v_{1}$, $v_{2}$, and $v_{5}$ are all step vertices in the final level.

Next, we give a non-standard, simple and  explicit definition of genericity of linkages for the properties studied in this paper as opposed to the more standard genericity of realizations.

\begin{definition}[Genericity]
\label{def:generic}
A 1-dof tree-decomposable linkage is \emph{generic} if no edge-length is zero, all edge-lengths are distinct, and at most one pair of adjacent edges is collinear in any realization. 
As noted in Section \ref{sec:Basic-properties}, we assume all clusters are minimally globally rigid, i.e. they generically have a unique realization and deleting any edge destroys this property, and so the length of any nonedge in a cluster is fixed in any realization of the linkage.  
For the purpose of the above genericity conditions, we treat any nonedge in a cluster, each of whose endpoints are shared by at least two clusters, as an edge.  
\end{definition}

\begin{remark}
    The reason for treating some nonedges in clusters as edges for the purpose of genericity is that, along some edge-length preserving continuous motion path of realizations, the local orientation of some triple of points may change.  
    When this change occurs, the local orientation becomes zero.  
    If all clusters are single edges, then some pair of adjacent edges becomes collinear at this point.  
    Otherwise, it may be that no pair of adjacent edges becomes collinear, but some pair of adjacent nonedges within clusters becomes collinear.  
    Our definition of generic requires that at most one pair of adjacent pairs of vertices -- which may be edges, nonedges within clusters, or a mix of the two -- is collinear in any realization.  
    Additionally, the issue of coincident points in Figure \ref{F:equal-distance}, discussed at the end of Section \ref{sec:Basic-properties}, may still occur if the edges $(v_7,v_5)$ and $(v_7,v_6)$ are instead nonedges within clusters.
\end{remark}

Next we define an important subclass of 1-dof tre-decomposable graphs, the subject of this paper.

\begin{definition}[Low Cayley algebraic complexity]
    \label{def:lowalg}
    A graph $G$ has low Cayley algebraic complexity if it is 1-dof tree-decomposable and there is a base nonedge $f$ s.t.  Cayley configuration space of every linkage of $G$ over $f$  has QRS interval endpoints.  \emph{low Cayley complexity (LCC)} has a slightly more restrictive definition given in Section \ref{sec:Combinatorial-interpretation-of}, but is conjectured to be equivalent \cite{bib:Owen02}.  
    An \emph{LCC linkage} is one with an underlying LCC graph.
\end{definition}

\begin{remark}
   A 1-dof tree-decomposable graph that has some base nonedge with strictly fewer than six clusters is \emph{trivially LCC}.  
\end{remark}

\subsubsection{Model of Computation}
Our complexity measures are based on a model of computation that uses exact representation
of numbers in any quadratic extension field of the rational numbers. In other words,
we assume that all arithmetic operations, over extraction of square roots and comparison are
constant time, exact operations. This model of computation is not as strong as the real RAM
model that is normally used in computational geometry, that permits exact representation
of arbitrary algebraic numbers \cite{bib:loos1983computing}. Issues in exact geometric computation such as efficient
and robust implementation of such a representation, for example using interval arithmetic,
are beyond the scope of this manuscript.

\subsection{Contributions}
\label{sub:contributions}

Theorem \ref{the:twobase}, stated below and proved in Section \ref{sec:Characterizing-parameter-choices:}, demonstrates that LCC is a robust property: if it holds for ( some base nonedge of ) a 1-dof tree-decomposable graph, then it holds for  all base nonedges of the graph. 

\begin{theorem}[Robustness of LCC] \label{the:twobase} 
A  graph is LCC on either all base nonedges or on none of them.
\end{theorem}

The above result relies on a complete combinatorial characterization of   LCC graphs stated in Theorem \ref{the:four-cycle} below and proved in Section \ref{sub:Characterizing-general}.  
A \emph{four-cycle} is four graphs such that each pair of graphs shares exactly one vertex and all four shared vertices are distinct.  
Any two of these graphs that share a vertex are said to be adjacent.  
Any pair of vertices that are not both contained in one of the graphs is a \emph{diagonal pair} if both vertices are shared, and a \emph{chordal pair} otherwise.

\begin{theorem}[Four-cycle Theorem]
\label{the:four-cycle}
    A 1-dof tree-decomposable graph with at least six clusters is nontrivially LCC if and only if it has a construction from some base nonedge such that, 
    for each construction step $v_{k}\triangleleft(u_k,w_k)$ with $v_k$ in $L_{2}$ or higher levels, $(u_k,w_k)$ is contained in an adjacent pair of clusters in some four-cycle of clusters.  
\end{theorem}

\begin{remark}
    If in addition $G$ is 1-path, then the construction has further structure (see Theorem \ref{thm:1-path-4-cycle} in Section \ref{sub:Characterizing-general}) used in the following results: Observation \ref{obs:GeneralTriCounter} and Theorem \ref{obs:k-path}.
\end{remark}

Theorem \ref{the:TriangleFreeCase} and Observation \ref{obs:GeneralTriCounter}, stated below and proved in Section \ref{sub:Forbidden-minor-characterization}, show that LCC is equivalent to planarity for a special class of 1-dof tree-decomposable graphs, and outside this class no such finite forbidden minor characterization exists.  
A graph is \emph{$K_3$-free} if it has no $K_3$ subgraph.  

\begin{theorem}[Equivalence of 1-path LCC to planarity]
\label{the:TriangleFreeCase} 
    A 1-path $K_3$-free 1-dof tree-decomposable graph is LCC if and only if it is planar. 
\end{theorem}

\begin{figure}[h]
\psfrag{1}{$v_{1}$} \psfrag{2}{$v_{2}$} \psfrag{3}{$v_{3}$}
\psfrag{4}{$v_{4}$} \psfrag{5}{$v_{5}$} \psfrag{6}{$v_{6}$}
\psfrag{7}{$v_{7}$} \psfrag{d13}{} \psfrag{d23}{} \psfrag{d14}{}
\psfrag{d24}{} \psfrag{d35}{} \psfrag{d45}{} 

\begin{centering}
\includegraphics[width=0.6 \textwidth]{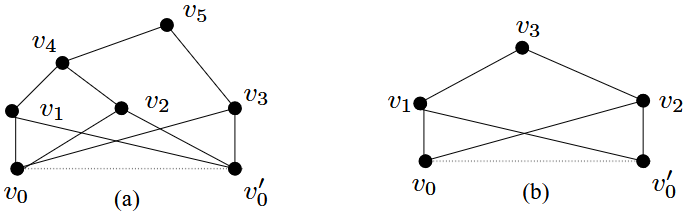} 
\par\end{centering}

\caption{Both (a) and (b) are $K_3$-free 1-dof tree-decomposable graphs.
(b) is LCC while (a) is not.  
See the discussion below.  }

\label{F:t-free} 
\end{figure}

Note that any tree-decomposable graph with more than two vertices contains a $K_3$ subgraph.  
Hence, every cluster of a $K_3$-free 1-dof tree-decomposable graph is a single edge.  
See Figure \ref{F:t-free}.

\begin{remark}
$K_3$-free 1-dof tree-decomposable graphs are related to Henneberg-I graphs \cite{bib:henneberg}.  
A Henneberg-I graph is a graph that can be constructed from a single edge be repeatedly adding a single vertex connected to exactly two existing vertices.  
A $K_3$-free 1-dof tree-decomposable graph can be obtained from a Henneberg-I graph by deleting the starting edge in its construction.  
\end{remark}

\begin{observation}[Limits of Theorem \ref{the:TriangleFreeCase}]
    \label{obs:GeneralTriCounter} 
    For any graph $H$, there 1-dof tree-decomposable graph $G$ that has exactly two of the following three properties and that has $H$ as a minor ($H$ can be obtained from both $G$ and $G'$ via sequences of edge deletions and contractions): (i) $G$ is LCC, (ii) $G$ is 1-path, and (iii) $G$ is $K_3$-free.  
\end{observation}

Theorem \ref{obs:k-path}, stated below and proved in Section \ref{sec:Find-Cayley-configuration},  describes, for any LCC   linkage, a \emph{minimal} $T$-oriented Cayley configuration spaces over any base nonedge, formally defined in Section \ref{sec:Find-Cayley-configuration}.  
The theorem further shows that this space is simple and easily computed.

\begin{theorem}[Fixed minimal realization type]
    \label{obs:k-path}
    For any LCC linkage with $n$ vertices, each of its minimal $T$-oriented Cayley configuration spaces over any base nonedge is (at most) a  single interval  whose endpoints can be computed in $O(n^2)$ time.  
\end{theorem}

The following observation, proved in Appendix \ref{sec:Exponential}, demonstrates minimality of the above $T$-oriented realization type.

\begin{observation}[Minimality of minimal realization type]
\label{obs: no-orientation}
For an LCC  linkage with $n$ vertices, if only a proper subset of the minimal $T$-orientations, such as the forward realization type, is specified  over some base nonedge,
then the corresponding oriented Cayley configuration space could contain exponentially many non-empty disjoint intervals in $n$.  
In fact, even deciding whether this oriented Cayley configuration space is non-empty does not have a polynomial time algorithm unless P = NP.  
In other words, listing all the interval (endpoints) could take exponential time even if the time complexity is measured w.r.t. output size, i.e. the number of nonempty intervals.
\end{observation}

Theorem \ref{theo:path}, stated below and proved in Section \ref{sec:cont-path}, is about finding paths between two realizations of a generic LCC linkage, and expressing the time to compute such a path in terms of the number of interval endpoints of forward oriented Cayley configurations spaces that it contains.  

\begin{theorem}[Continuous motion path] \label{theo:path}
For any generic LCC linkage and any base nonedge $f$, there exist at most two edge-length preserving continuous motion paths between any two realizations of the linkage, and such a path can be found (provided one exists) in time linear in the number of interval endpoints of the forward oriented Cayley configuration spaces over $f$ that the path contains.  
Furthermore, if the realizations belong to the same minimal $T$-oriented Cayley configuration space, then such a path between them can be found in constant time.  
\end{theorem}

\begin{remark}
    For any generic LCC linkage, to obtain a continuous path between two of its Cayley configurations where their forward realization types are unspecified, 
    we run the algorithm given by Theorem \ref{theo:path} for each candidate forward realization type of the starting Cayley configuration and each candidate forward realization type of the target Cayley configuration.
    For each pair of starting and target realizations there are at most two paths between them, but the number of such pairs could be exponential in the size of the linkage. 
\end{remark}

\begin{remark}
    Implementation of the algorithms discussed in this paper for finding Cayley configuration spaces is part of our  CayMos software, whose architecture is described in \cite{bib:caymos}. 
A different manuscript \cite{bib:beast} 
describes Cayley and Cartesian configuration space analysis and motion analysis 
of common and well-known mechanisms using CayMos.
\end{remark}

Finally, Theorem \ref{thm:parameterize}, stated below and proved in Section \ref{sec:ambient}, provides a means of obtaining a bijection between the space of 2D realizations of a generic LCC linkage $(G,\bar{l})$ and the Cayley configuration space over a particular nonedge set $F$, i.e., the set of length vectors attained by $F$ over all 2D realizations.  
The set $F$, called a \emph{minimal bijective Cayley vector} of $G$, is any set of nonedges that contains some base nonedge of $G$ and such that $G \cup F$ is minimally globally rigid.  
We give an algorithm in Section \ref{sec:ambient} to construct a family of minimal bijective Cayley vectors of size $n+1$, where $n$ is the number of final level step vertices for some base nonedge.  
As noted in Section \ref{sec:Basic-properties}, we assume all clusters of $G$ are minimally globally rigid, and so Theorem \ref{thm:parameterize} makes the reasonable assumption that wlog all clusters sharing exactly two vertices with the rest of the graph are treated as single edges.  

\begin{theorem}[Bijectivity of representation] 
\label{thm:parameterize}
Consider a 1-dof tree-decomposable linkage with graph $G$ such that each of its clusters is minimally globally rigid and each cluster that shares exactly two vertices with the rest of the graph is a single edge, and let $F$ be a minimal bijective Cayley vector of $G$ containing some base nonedge $f$ as constructed in Section \ref{sec:ambient}.  
There exists an arbitrarily small perturbation of the edge-length vector such that there is a bijection between the space of all 2D realizations of the linkage and its Cayley configuration space over $F$, and the latter space is a curve in $(n+1)$-dimensions, where $n$ is the number of final level step vertices for $f$.  
\end{theorem}

\subsection{Relation to Other Works}
\label{sub:relation}
\begin{figure}[hbtp]
	\centering
	\includegraphics[width=8cm]{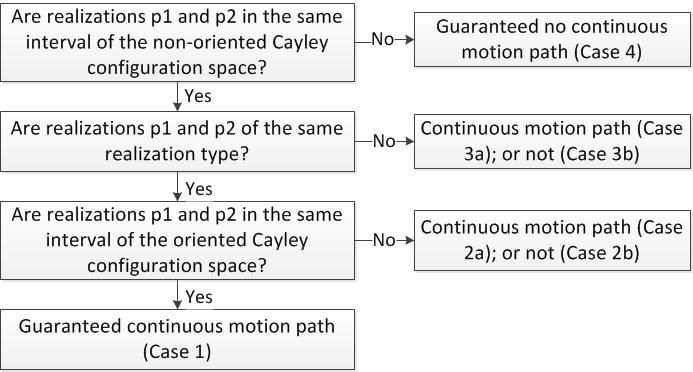}
	\caption{Complete case analysis of continuous motion paths between two realizations $p_1$ and $p_2$. 
    See the discussion below.  
	}
	\label{fig:diagram}
	\end{figure}

As mentioned earlier, relevant previous work primarily dealt with continuous motion paths between realizations.
Figure \ref{fig:diagram} summarizes the different cases when 
	determining existence of a continuous motion path between two realizations. 
	There are two cases (2 and 3) where there may or may not exist a continuous motion path (a and b).
   We know that the Cayley configuration is not always a bijective representation of the realization space
	Specifically, a non-oriented Cayley interval, being a union of multiple oriented Cayley intervals, 
	could correspond to multiple connected components of the realization space, 
	as in Figure \ref{F:interval}. 
	Although an oriented Cayley interval corresponds to a unique connected component, 
	the mapping is not bijective, since that same connected component could contain more than one oriented interval.	
	Previous algorithms and software, except for \cite{bib:hidalgo2011reachability},
	generate continuous motion within a specified Cayley interval, 
	or multiple segments of continuous motion, 
	each corresponding to different oriented Cayley interval with the same realization type. 
	Thus they cannot consistently distinguish Case 2a from Case 2b, 
	or Case 3a from Case 3b.
	The algorithms in \cite{bib:hidalgo2011reachability} can distinguish between these four types. 
	However, it deals with general tree-decomposable linkages, relies on exhaustive searching and could have exponential time complexity.

\begin{figure}[hbtp]
    \centering
	\includegraphics[width= .6\linewidth]{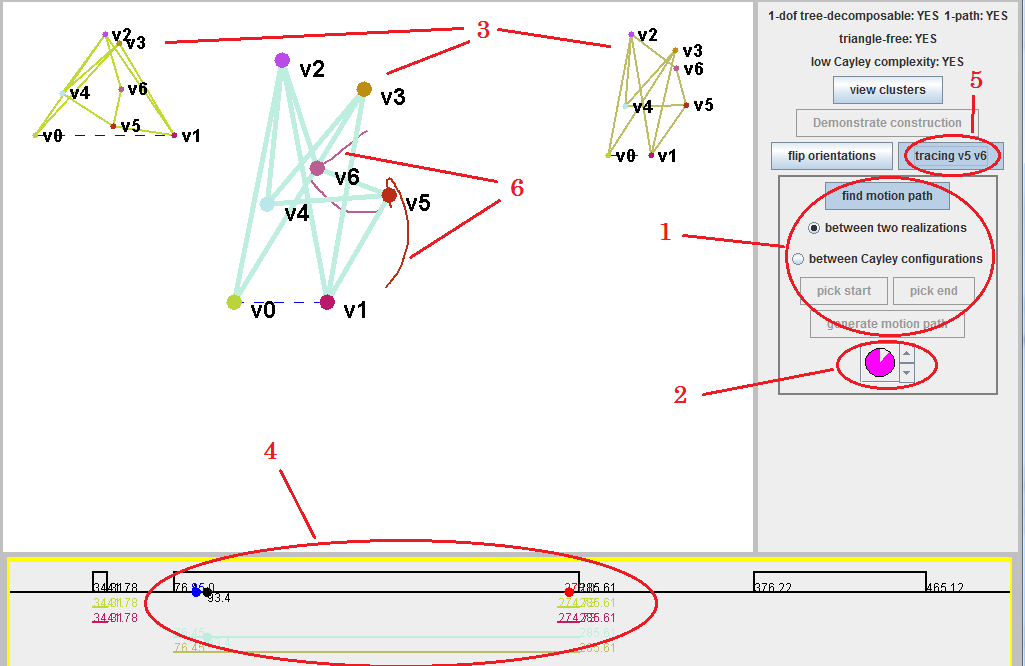}

\caption{Finding a continuous motion path using  our CayMos software \cite{bib:caymos} between two realizations. 
(A) The start and end realizations.
(B) The current realization, moving as the user traces the continuous motion path. 
(C) The 3D projection of the current connected component in the motion space. 
(D) The intervals of the oriented Cayley configuration spaces encountered along the path. 
See the discussion below.  }
\label{F:caymos_path}
\end{figure}

Finally, the papers \cite{bib:beast,bib:caymos}	have implemented the algorithms of this paper in an opensource software CayMoS  github/geoplexity/caymos, shown that the LCC class contains many common linkages including the strandbeest \cite{strandbeestStrandbeest}, and analyzed the realization spaces and motions of these linkages using CayMos.
 CayMos architecture is described in \cite{bib:caymos}.  
 Demonstration videos are available with \cite{bib:beast}, or upon request, see also the screen-shot in Figure \ref{F:caymos_path}.

 \section{Basic properties of 1-dof tree-decomposable  linkages and their Cayley Configuration Spaces}
 \label{sec:Basic-properties}

Recalling the notation for the construction of 1-dof tree-decomposable graphs from a base nonedge $f$ in Section \ref{sec:contributions}, at the $k^{th}$ step of the construction, vertices $u_k$, $v_k$ are called the {\emph{base pair of vertices}} at Step $k$. 
Since $T_{1}$ and $T_{2}$ are maximal clusters, it follows that $u_k$ and $v_k$ are not both contained in some cluster in $G_f(k-1)$, the graph after the $(k-1)^{st}$  construction step.
Hence, they are not connected by an edge and they lie in distinct clusters $T_{u_k}$ and $T_{v_k}$ in $G_{f}(k-1)$, called the \emph{$k^{th}$ base pair of clusters}.  
The construction steps can be arbitrarily ordered consistent with the levels defined in Section \ref{sec:contributions}, as long as, at the $k^{th}$ construction step, $G_{f}(k-1)$ contains $u_k$ and $v_k$.  
If a vertex $v$ is shared by $m$ distinct clusters, we say {$cdeg(v)=m$}.  
The following is a direct consequence of the results in \cite{bib:FudHo97proof}.  

\begin{observation}[Clusters are unique]
     The set of clusters of a 1-dof tree-decomposable graph $G$ is unique, and hence constructions of $G$ from any two base nonedges have the same number of construction steps.  
 \end{observation}

 For example, refer to the 1-dof tree-decomposable graph $G$ and base nonedge $f = (v_0,v'_0)$ in Figure \ref{F:construction}(a).  
Two of the construction steps of $G$ are 
$v_{1}\triangleleft(v_{0}\in T_{1},v_{0}'\in T_{2})$ and $v_{7}\triangleleft(v_{5},v_{6})$, which yield graphs $G_f(1)$ and $G_f(7)$ respectively. 
 Note that $cdeg(v_7)=3$.  
 
 The  level sets of the clusters and vertices of a 1-dof tree-decomposable graph  provide a better understanding the constructions from a base nonedge.

\begin{observation}[Final level vertices tthat are unique]
The  final level vertices  are the same for all base nonedges whose endpoints lie outside the final level clusters.
 More precisely, for  a final level vertex $v$ shared by two clusters $T_1$ and $T_2$, 
let $G'=G\setminus(T_{1}\cup T_{2})$, $T_{1}\cap G'=\{u\}$, $T_{2}\cap G'=\{w\}$. 
The construction step 
$v\triangleleft(u\in T_{1},w\in T_{2})$ can be taken as the last step for construction of 
$G_f$ from any base nonedge that is presented in $G'$. 
\end{observation}

\smallskip
 \noindent\textbf{Note.} Throughout this paper we assume fixed realizations of the clusters of $G$ when constructing a realization of $(G,\ell)$,  i.e., we assume all clusters are minimally globally rigid.  
 Realizations of the clusters have no bearing on the results in this paper.   
 Moreover,   all  linkages are generic (see Definition \ref{def:generic}, avoiding special lengths, leading to the following observation.

\begin{observation}[Bijection for forward realization types]
 \label{obs:1-realization}
 For any generic 1-dof tree-decomposable linkage, each point in a forward oriented Cayley configuration space $X$ corresponds to exactly one 2D realization of this linkage that agrees with the forward realization type that defines $X$.
\end{observation}

A generic linkage $(G,\bar{l})$ with a 1-dof tree-decomposable underlying graph $G$   has one degree of freedom,
and a  Cayley configuration space with parameter $f$.
Hence, 
  for a given length $l_f$ of a base nonedge $f$, a  realization $p$ of a 1-dof tree-decomposable linkage $(G,\ell)$ can be computed, if one exists, via a ruler-and-compass (QRS) using a construction of $G$ from a base nonedge $f$.  
 The clusters in the $k^{th}$ construction step $w \triangleleft (u,v)$ from $f$ can be placed by computing $p_w$, which, if it exists, is one of two intersection points of two circles centered at $p_u$  and $p_v$, whose radii are known from the fixed realizations of the base pair of clusters.
The two intersection points correspond to different local orientations of $(u,v,w)$.  

\begin{figure}[h]
	\psfrag{1}{$v_{1}$} \psfrag{2}{$v_{2}$} \psfrag{3}{$v_{3}$}
	\psfrag{0}{$v_{0}$} \psfrag{0'}{$v_{0}'$} 
	
	\begin{centering}
	\includegraphics[width=0.8\textwidth]{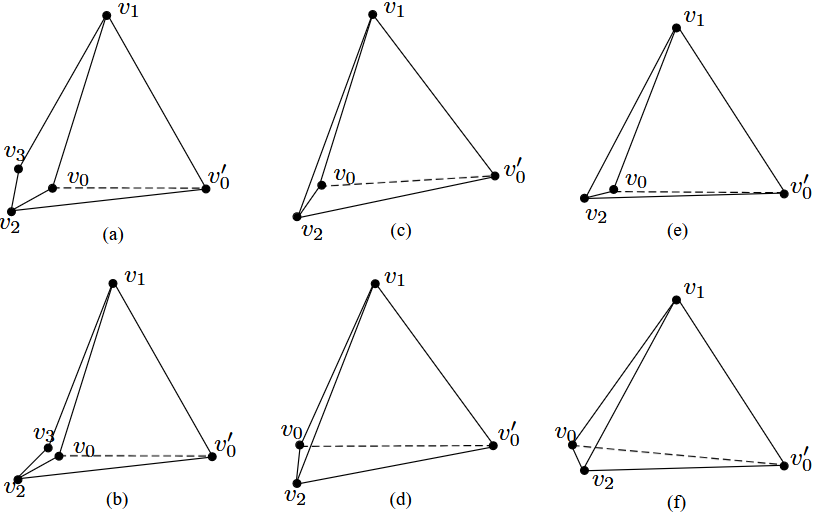}
	\par\end{centering}
	
	\caption{ (a) (b): Two choices exist for local orientation of
	construction step $v_{3}\triangleleft(v_{1},v_{2})$ from base nonedge $f=(v_{0},v_{0}')$. 
	(c) (d): realizations for extreme linkage $(\hat{G}_{f}(3),\bar{l}^{\max})$ (defined in Section \ref{sec:Combinatorial-interpretation-of}) 
	(e) (f): realizations for extreme linkage $(\hat{G}_{f}(3),\bar{l}^{\min})$ (defined in Section \ref{sec:Combinatorial-interpretation-of}).  
    See the discussion below and Section \ref{sec:Find-Cayley-configuration}.}

	\label{F:direction} 
\end{figure}

For example, refer to Figure \ref{F:direction} (a) and (b).  
The graph is 1-dof tree-decomposable with base nonedge $f=(v_{0},v_{0}')$.  
Two possible local orientations corresponding to the construction step $v_{3}\triangleleft(v_{1},v_{2})$ are shown, which yield different forward realization types $\sigma$ and $\sigma'$:
(a) has $\sigma_3 < 0$, 
while (b) has $ \sigma'_3 > 0$ 

Hence, the time to compute a realization $p$, if it exists, could be exponential in the number of construction steps, but  is linear in this number if a forward realization type is given.  

\begin{figure}[h]
	\psfrag{1}{$v_{0}$} \psfrag{2}{$v_{0}'$} \psfrag{3}{$v_{1}$}
	\psfrag{4}{$v_{2}$} \psfrag{5}{$v_{3}$} \psfrag{6}{$v_{4}$}
	\psfrag{7}{$v_{5}$} \psfrag{8}{$v_{6}$} \psfrag{9}{$v_{7}$}
	\psfrag{10}{$v_{8}$} 
	
	\begin{centering}
	\includegraphics[width=0.8\textwidth]{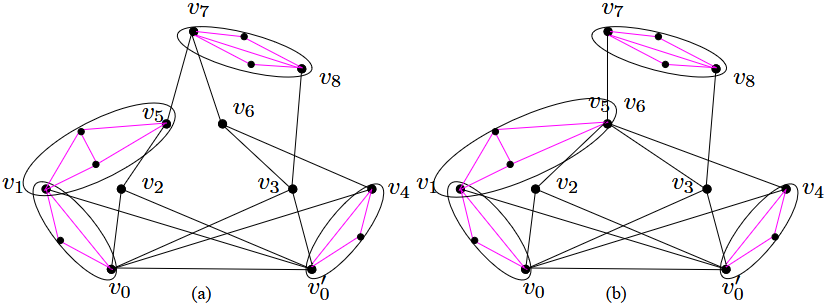}
	\par\end{centering}
	
	\caption{Example showing the importance of genericity: when $p(v_{5})$ and $p(v_{6})$ are coincident, the length of $(v_{5},v_{8})$
	is not a function of the length of $(v_{0},v_{0}')$.  
    See the discussion below and Section \ref{sub:definitions}.}
	
	\label{F:equal-distance} 
\end{figure}

Moreover, as demonstrated in Figure~\ref{F:equal-distance}, the genericity assumption is crucial:
for any vertex $v$ of $G$, the point $p_v$ is not unique only if, for $v$'s construction step $v \triangleleft (u,w)$,
$p_u$ and $p_w$ are coincident and $\bar{l}(v,u) = \bar{l}(v,w)$.

\subsection{LCC and Extreme graphs}
\label{sec:Combinatorial-interpretation-of}

Here we define LCC  precisely (differentiating from low Cayley algebraic complexity) by examining the structure of forward oriented Cayley configuration spaces in Theorem \ref{lem:algebraic}, below.  
First, we define \emph{extreme} graphs and linkages.  

\begin{definition}[Extreme graphs and linkages]
\label{def:extreme-graph}
Consider the construction of a 1-dof tree-decomposable graph $G$ from a base nonedge $f$.  
The \emph{$k^{th}$ extreme graph} $\hat{G}^f(k)$ is obtained from $G_{f}(k-1)$ by adding the $k^{th}$ base pair of vertices $(u_k,w_k)$ as an edge, which we call the \emph{extreme edge} of $\hat{G}_{f}(k)$, 
and the \emph{$k^{th}$ extreme nonedge} for $f$.  

Consider the $k^{th}$ construction step $v_k \triangleleft (u_k,w_k)$.  
For any linkage $(G,\bar{l})$, the \emph{$k^{th}$ extreme linkages} are $(\hat{G}_{f}(k),\bar{l}^{\min})$
and $(\hat{G}_{f}(k),\bar{l}^{\max})$, where $\bar{l}^{\min}$ and $\bar{l}^{\max}$ represent both agree with $\ell$ on the edgees of $G_f(k-1)$, 
$\bar{l}^{\min}(u_k,w_k):=|\bar{l}(u,v_{k})-\bar{l}(v_{k},w)|$, and
$\bar{l}^{\max}(u_k,w_k):=\bar{l}(u,v_{k})+\bar{l}(v_{k},w)$. 
\end{definition}

\begin{remark}
It is easy to verify using Laman's theorem \cite{bib:Laman70} that any extreme graph of a 1-dof tree-decomposable graph is minimally rigid.  
In realizations of $(\hat{G}_{f}(k),\bar{l}^{\min})$ and $(\hat{G}_{f}(k),\bar{l}^{\max})$, the local orientation of $(u_k,w_k,v_k)$ is $0$.
These realizations are sometimes called \emph{unyielding} realizations.
\end{remark}

For example, the graph $\hat{G}_f(7)$ in Figure \ref{F:construction}(b) is $G$'s extreme graph corresponding to Construction Step $7$ from $f$, 
obtained by adding the edge $(v_5,v_6)$ to $G_f(6)$.  


\begin{theorem}[Structure of Cayley configuration space]\label{lem:algebraic}
 For any generic 1-dof tree-decomposable linkage $(G,\bar{l})$ with base nonedge $f$, 
the following hold:

 \begin{enumerate}
 	\item The (resp. forward oriented) Cayley configuration space over $f$ is either empty or the union of a set of disjoint closed real intervals. 
 	
 	\item Any interval endpoint in the (resp. oriented) Cayley configuration space corresponds to the length of $f$ in a realization of some extreme linkage. 
 	
 	\item For any vertex $v$, $p_v$ is a continuous
 	 	function of $l_f$ on each closed interval of the oriented Cayley configuration space.
 	 	Consequently, for any  nonedge $(u,w)$, $l(u,w)$ is a continuous function of $l_f$ on each closed interval of the oriented Cayley configuration space.
 \end{enumerate}
 
\end{theorem}

The proof of Theorem \ref{lem:algebraic} follows from elementary algebraic, and is given in Appendix \ref{sec:Proof-for-lemma} for completeness.

While Theorem \ref{lem:algebraic} (2) states that every endpoint of the unoriented Cayley configuration space
corresponds to the length of $f$ in a realization of some extreme linkage, 
the converse is not true. 
As an example, refer to Figure \ref{F:interval}. 
Realization (D) is an extreme linkage, 
but it is not an endpoint in the unoriented Cayley configuration space.
However, Theorem \ref{thm:interval-endpoints}, below, shows that the converse is true for forward oriented Cayley configuration spaces. 

\begin{theorem}[Characterization of forward oriented interval endpoints]
    \label{thm:interval-endpoints}
    Consider any forward oriented Cayley configuration space $X$ of a generic 1-dof tree-decomposable linkage $(G,\bar{l})$ over a base nonedge $f$.
    A point in $X$ corresponds to the length of $f$ in a realization of some extreme linkage of $(G,\bar{l})$ if and only if it is an endpoint of some interval in $X$.  
\end{theorem}

\begin{proof}
    The converse direction is Part (2) of Theorem \ref{lem:algebraic}, which is proved in Appendix \ref{sec:Proof-for-lemma}.  
    For the forward direction, assume to the contrary that some interior point $x \in X$ corresponds to the length of $f$ in a realization of some extreme linkage of $(G,\bar{l})$.  
    Wlog, let this extreme linkage be $(G_f(k),\bar{l}^{max})$ and let its extreme edge be $e = (u,v)$.  
    Recall from Observation \ref{obs:1-realization} in Section \ref{sec:Basic-properties} that there is a bijection between $X$ and realizations of $(G,\bar{l})$ that agree with the forward realization type that defines $X$.  
    Hence, for any point $y \in X$, let $p_y$ be its corresponding realization.  
    Also, let $p_y(e)$ be the length of $e$ in $p_y$.  
    Lastly, let $I$ be the interval in $X$ that contains $x$.  
    Then, in some sufficiently small neighborhood of $x$ in $I$, there exist points $y$ and $z$ such that $y < x < z$, $p_y \neq p_z$, $p_x(e) = \bar{l}^{max}(e)$, and $p_y(e),p_z(e) < p_x(e)$.  
    These facts imply that either (i) the triple $(u,v,w)$, where $w$ is the $k^{th}$ step vertex, has distinct local orientations in $p_y$ and $p_z$ or (ii) Statement (i) is false and so some other triple $(u',v',w')$ of vertices has distinct local orientations in $p_y$ and $p_z$.  
    If Statement (i) is true, then $p_y$ and $p_z$ do not have the same forward realization type, which we know is not possible since $y$ and $z$ are contained in $X$.  
    If (ii) is true, then all vertices along some path containing $u'$, $v'$, and $w'$ are collinear in $p_x$.  
    Since $u$, $v$, and $w$ are collinear in $p_x$, this contradicts the genericity of $(G,\bar{l})$.  
    Thus, $x$ must be an endpoint of $I$, and so the proof is complete.  
\end{proof}

\begin{remark}
    While the proof of the forward direction of Theorem \ref{thm:interval-endpoints} requires genericity, the proof of the converse direction does not.  
\end{remark}


\begin{observation}[ELR algorithm]
    \label{obs:NP-hard}
    Theorem \ref{lem:algebraic} gives a straightforward algorithm  called {\emph{ELR (extreme linkage realization)}} 
    to obtain the (forward oriented) Cayley configuration space for a generic 1-dof tree-decomposable linkage $(G, \bar{l})$, which could take time exponential in the number of intervals it contains.  
\end{observation}

\begin{proof}
The ELR algorithm works by realizing all the extreme linkages for $f$ consistent with each forward realization type.  
Even if the extreme graphs are QRS, even determining the existence of, let alone finding, extreme linkage realizations is NP-hard.  
Since the extreme graphs may not be QRS, realizing each extreme linkage can take time exponential in $|V|$ 
(requiring the solution of a general multi-variable system of quadratic equations).
Additionally, the overall time complexity could be exponential in the actual number of intervals in the Cayley configuration space, 
since many candidate endpoints generated during this procedure could finally lead to dead ends. 
The detailed version of this algorithm is in Appendix \ref{sec:not-low}. 
\end{proof}

\begin{remark}
    From a complexity point of view, the observation is unsurprising because the problem of determining the existence of a realization of a tree-decomposable linkage, which
is NP-complete by early results \cite{bib:saxe79}, can be reduced to the decision version of our problem: deciding whether the Cayley configuration space over $f$ is non-empty.  
Clearly, a realization exists if and only if the Cayley configuration space is not empty. 
Therefore, deciding whether the Cayley configuration space over $f$ is non-empty is NP-hard, and can take superpolynomial time unless $P = NP$.  
\end{remark}

\begin{definition}[Low Cayley complexity]
    \label{def:low}
    A 1-dof tree-decomposable $G$ is LCC on base nonedge $f$ if all extreme graphs of $G$ for $f$ are tree-decomposable. 
\end{definition}

\begin{figure}[h]
	\psfrag{1}{$v_{0}$} \psfrag{2}{$v_{0}'$} \psfrag{3}{$v_{1}$}
	\psfrag{4}{$v_{2}$} \psfrag{5}{$v_{3}$} \psfrag{6}{$v_{4}$}
	\psfrag{7}{$v_{5}$}
	
	\begin{centering}
	\includegraphics[width=0.7\textwidth]{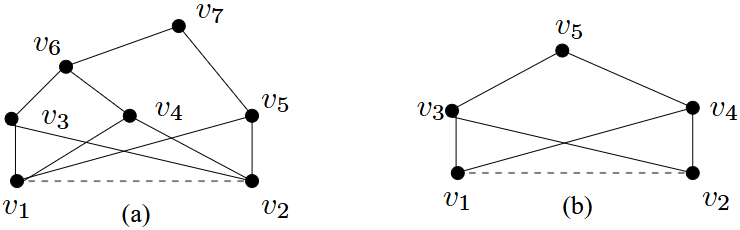}
	\par\end{centering}
	
	\caption{The graph in (b) is LCC while the graph in (a) is not.  
    See the discussion below.  }
	
	\label{F:not_low} 
\end{figure}

Observe that a LCC graph has low Cayley algebraic complexity (see Definition \ref{def:lowalg} in Section \ref{sub:definitions}).  
For example, in Figure \ref{F:1-path}, (a) and (c) are LCC on $(v_0,v'_0)$, while (b) and (d) are not, 
since both of them have the extreme graph $\hat{G}_{f}(4)$ that is not tree-decomposable. 
As another example, refer to Figure \ref{F:not_low}.  
The graph in (b) is LCC on $(v_0,v_0')$, while the graph in (a) is not, since the extreme graph corresponding to the construction step whose step vertex is $v_5$ is not tree-decomposable.  
We can also verify that the 1-dof tree-decomposable graphs in Figure \ref{F:intro}, Figure \ref{F:construction} and Figure \ref{F:direction} are LCC on the given base nonedge, while the graph in Figure \ref{F:equal-distance} is not.

\section{Theorem \ref{the:twobase}: Choice of base nonedge immaterial for LCC graphs}
\label{sec:Characterizing-parameter-choices:}

Here we prove Theorem \ref{the:twobase}, which shows that a 1-dof tree-decomposable $G$ is LCC on all or none of its base nonedges.  
The proof is by induction on the number $n$ of construction steps in the construction $G$ from any base nonedge $f$.  
The base case, where $n \leq 2$, follows from Definition \ref{def:low}, and the inductive hypothesis tells us that $G_f(k)$ is LCC on all or none of its base nonedges, for any $k \geq 2$.  
From this, we will get that $G$ is LCC on all or none of its base nonedges in $G_f(k)$.  
Finally, for any construction of $G$ from a base nonedge $f'$ with at least one endpoint not in $G_f(k)$, we will show how to choose the first two construction steps such that the clusters they add contain some base nonedge $f''$ in $G_f(k)$ for which $G$ is LCC on $f'$ if and only if it is LCC on $f''$.  
This completes the proof.  


\begin{proof}[Proof of Theorem \ref{the:twobase}]
Let $G$ be a $1$-dof tree-decomposable graph. 
We will show by induction on the number $n$ of construction steps in the construction of $G$ from any base nonedge $f$ that $G$ is LCC on all base nonedges or none of them.  
When $n \leq 2$, the theorem follows easily from Definition \ref{def:low}.  
Assume the theorem holds when $n = k$, for any $k \geq 2$, and we will prove it when $n = k + 1$.  
By the inductive hypothesis, $G_f(k)$ is LCC on all base nonedges or none of them.  
Let $v \triangleleft (u \in T_1,w \in T_2)$ be the last construction step.  
Since $T_1$ and $T_2$ each share exactly one vertex with $G_f(k)$, any construction of $G$ from a base nonedge must perform $v \triangleleft (u \in T_1,w \in T_2)$, and we can clearly perform this construction step last.  
Hence, $G$ is LCC on all base nonedges in $G_f(k)$ or none of them.  
Therefore, it suffices to show $G$ is LCC on any base nonedge in $G_f(k)$ if and only if it is LCC on any base nonedge $f'$ that has at least one endpoint $x$ in $(T_1 \cup T_2) \setminus \{u,w\}$.  

Let the first two construction steps from $f'$ be $v_1 \triangleleft (u_1,w_1)$ followed by $v_2 \triangleleft (u_2,w_2)$.  
Note that $cdeg(x)$ is $2$ if $x = v$, and $1$ otherwise.  
If $v_1 \triangleleft (u_1,w_1)$ does not add $T_1$ or $T_2$, then $cdeg(x)$ is strictly greater than $2$ if $x = v$, and strictly greater than $1$ otherwise.  
Combining these facts shows that $v_1 \triangleleft (u_1,w_1)$ adds either $T_1$ or $T_2$.  
Let $f''$ be $(v_1,v_2)$ if $(v_1,v_2)$ is contained in $G_f(m-1)$, and let $f''$ be $(u_2,w_2)$ otherwise.  
In the former case, note that $G$ can be constructed from $f''$ by choosing the first two construction steps from $f''$ to add the same clusters as the first two construction steps from $f'$, and then proceeding identically to the construction from $f'$.  
Therefore, $G$ is LCC on $f'$ if and only if it is LCC on $f''$, as desired.  

Finally, in the latter case, a similar argument shows that $G$ is LCC on $f'$ if and only if it is LCC on $f''$.  
Additionally, since $cdeg(v_1)$ and $cdeg(v_2)$ are both at least $2$ and $v_1$ and $v_2$ are distinct, wlog we get that $v_1 = v$ and $v_2$ is contained in $G_f(k)$.  
Hence, $v_1 \triangleleft (u_1,w_1)$ adds both $T_1$ and $T_2$, and so $f''$ is contained in $G_f(k)$.  
Thus, combining these facts completes the proof.  
\end{proof}




\section{Theorem \ref{the:four-cycle}: A combinatorial characterization of LCC (four-cycles)
and an efficient recognition algorithm}
\label{sub:Characterizing-general}

In this section we prove Theorem \ref{the:four-cycle}, the ``Four-cycle Theorem''.  

\begin{figure}[h]
\psfrag{T1}{$T_{1}$} \psfrag{T2}{$T_{2}$}\psfrag{T3}{$T_{3}$}\psfrag{T4}{$T_{4}$}\psfrag{T5}{$T_{5}$}\psfrag{T6}{$T_{6}$}

\psfrag{p1}{$p_{1}$}\psfrag{p2}{$p_{2}$}\psfrag{p3}{$p_{3}$}\psfrag{p4}{$p_{4}$}

\psfrag{v1}{$v_{0}$} \psfrag{v2}{$v_{0}'$} \psfrag{vk}{$v_{k}$}
\psfrag{G_k-1}{$G_{f}(k-1)$} 

\psfrag{f}{$f$} \psfrag{u}{$u_{k}$} \psfrag{w}{$w_{k}$} 

\begin{centering}
\includegraphics[width=.8\linewidth]{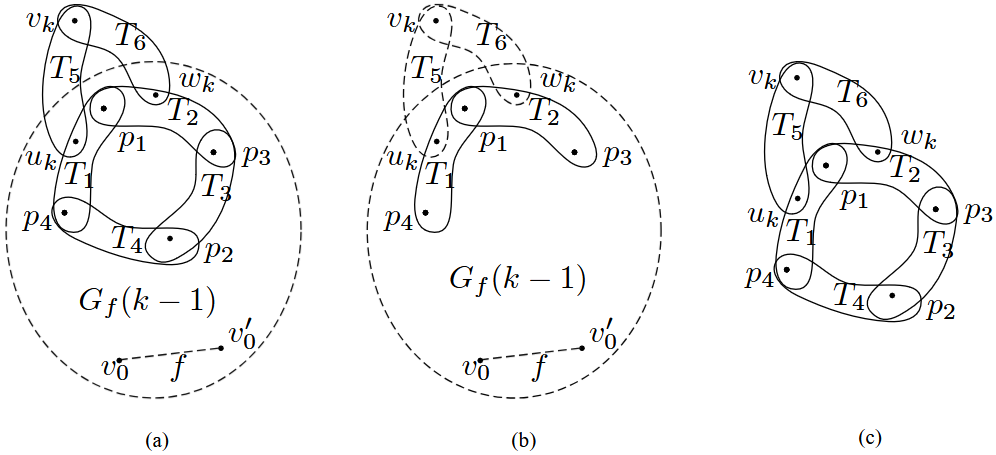} 
\par\end{centering}

\caption{Graphs in the proof of Theorem \ref{the:four-cycle}.  
See the discussion below.  }

\label{F:fourcycle} 
\end{figure}

\begin{proof}[Proof of Theorem \ref{the:four-cycle}]
Let $G$ be a 1-dof tree-decomposable graph with at least six clusters, and consider its construction from some base nonedge $f$.  
For the forward direction, assume $G$ is nontrivially LCC and let $v_{k}\triangleleft(u_{k},w_{k})$ be any construction step with $v_k$ in $L_2$ or higher levels. 
Then, the extreme graph $\hat{G}_{f}(k)$ is tree-decomposable, and so $\hat{G}_{f}(k)$ can be constructed from $(u_k,w_k)$. 
Note that this construction has at least two construction steps, and as in Figure \ref{F:fourcycle} (b), the clusters added by the first two construction steps form a four-cycle such that $(u_k,w_k)$ is contained in an adjacent pair of its clusters, as desired.  

Next, we prove the converse by induction on the number $n$ of construction steps in the construction of $G$ from $f$.  
Since $G$ has at least six clusters, we have $n \geq 3$.  
When $n=3$, $G$ is shown in Figure \ref{F:fourcycle} (c), and is clearly LCC.  
Assume the converse holds when $n = k-1$, for any $k \geq 4$, and we will prove it when $n=k$.  
Wlog, we can assume $f$ is contained in $G_f(k-1)$, which is LCC by the inductive hypothesis.  
It suffices to show that $\hat{G}_f(k)$ is tree-decomposable, which we do now.  

Consider the construction step $v_k \triangleleft (u_k,w_k)$.  
If $v_k$ is in $L_1$, then $\hat{G}_f(k)$ is clearly tree-decomposable
Otherwise, $v_k$ is in $L_2$ or higher levels, and so, by assumption, there exists a four-cycle $C$ of clusters $T_1$, $T_2$, $T_3$, and $T_4$ such that $T_1$ and $T_2$ are adjacent and contain $(u_k,w_k)$, as in Figure \ref{F:fourcycle} (a).  
Of the construction steps that add clusters in $C$, the last one, $v_m \triangleleft (u_m,w_m)$ where $m < k$, must add two of these clusters, wlog say $T_1$ and $T_2$.  
Otherwise, $G_f(m-1)$ contains the three clusters in $C$, the last cluster added by this construction step shares two vertices with $G_f(m-1)$, which violates the definition of a construction step.  

Finally, since $G_f(k-1)$ is LCC, we get that $\hat{G}_{f}(m)$ is tree-decomposable.  
Hence, $G_f(m)$ can be constructed from $(u_m,w_m)$, and consequently $G_f(k-1)$ can also be constructed from $(u_m,w_m)$.  
An argument similar to that above shows that some construction step in the construction of $G_f(k-1)$ from $(u_m,w_m)$ must add both $T_3$ and $T_4$.  
Therefore, we can perform this construction step first, followed immediately by $v_m \triangleleft (u_m,w_m)$.  
Since the above four-cycle, which contains $(u_m,w_m)$, can be constructed from $(u_k,w_k)$, the above facts show that $G_f(k-1)$ can be constructed from $(u_k,w_k)$.  
Thus, $\hat{G}_f(k)$ is tree-decomposable, as desired.  
\end{proof}

Theorem \ref{the:four-cycle} yields an algorithm, stated below, to verify whether a 1-dof tree-decomposable graph $G = (V,E)$ is LCC in $O(|V|^2)$-time. 
This is more efficient than the algorithm that follows from the definition of LCC, i.e., checking if all extreme graphs of $G$ are tree-decomposable, which takes $O(|V|^3)$-time (checking $O(|V|)$ extreme graphs, each taking $O(|V|^2)$-time using the existing algorithm in \cite{bib:FudHo97}). 
Our algorithm follows the construction of $G$ from $f$ and maintains
a list $L$ of adjacent cluster pairs that can contain the base pair of vertices for the next construction step. 

\medskip
\noindent \textbf{Algorithm (Recognizing LCC)}:
\begin{enumerate}
    \item Given a 1-dof tree-decomposable graph $G$, consider any construction of $G$ from any base nonedge $f$, and let all construction steps that add first level vertices be performed before all other construction steps.  
    Start with an empty set $L$ and add to it all pairs of first level clusters of $G$ that share at least one vertex.  

    \item For each remaining construction step $v_{k}\triangleleft(u_{k}\in T_k,w_{k}\in T_k')$, going from lowest to highest index, do the following:
    \begin{enumerate}[a.]
        \item Find the following two sets of clusters: $U=\{T_{u}|T_{u}\in G_{f}(k-1),u_{k}\in T_{u}\}$, $W=\{T_{w}|T_{w}\in G_{f}(k-1),w_{k}\in T_{w}\}$.

        \item Find all the pairs $(T_{u},T_{w})$ that are contained in both $U\times W$ and $L$; if no such pair exists, return that $G$ is not LCC.  

        \item For all pairs $(T_{u},T_{w})\in U\times W$ that share at least one vertex, add $(T_k,T_{u})$ and $(T_k',T_{w})$ to $L$.  
        Add $(T_k,T_k')$ to $L$.
    \end{enumerate}
    \item Return that $G$ is LCC.  
\end{enumerate}


Let $n$ be the total number of construction steps in the construction of $G$ from $f$. 
Step (1) and and Step (3) take $O(n^2)$-time and $O(1)$-time, respectively. 
For each iteration of Step (2), Step (a) takes $O(cdeg(u_k)+cdeg(w_k))$-time and Steps (b) and (c) each take $O(cdeg(u_k) \times cdeg(w_k))$-time. 
Since $cdeg(u_k)$ and $cdeg(w_k)$ are both $O(n)$, the overall time complexity is $O(n^2)$, which is $O(|V|^2)$. 

As noted in Section \ref{sec:contributions} and stated below in Theorem \ref{thm:1-path-4-cycle}, the four-cycle characterization of LCC graphs has more structure when the graphs are 1-path.  
The proof of this theorem requires Lemmas \ref{lem:1-path-lcc-base} and \ref{the:chain}, also below.  
Lemma \ref{lem:1-path-lcc-base} is proved using Lemma \ref{lem:1-path-2-level-1}, which we now state and prove.  

\begin{lemma}[1-Path LCC first level step vertices]
    \label{lem:1-path-2-level-1}
    Let $G$ be a 1-dof tree-decomposable graph with at least six clusters, and consider its construction from some base nonedge $f$.  
    If $G$ is LCC and has a 1-path construction from $f$, then $f$ has at most two first level step vertices.  
\end{lemma}

\begin{proof}
    Assume that $f$ has at least three first level step vertices.  
    If $G$ does not have a 1-path construction from $f$, then we are done.  
    Otherwise, order the construction steps for $f$ such that all first level step vertices are added before any other step vertices.  
    For any integer $i \geq 1$, let the final level step vertices of $G_f(i)$ be the final level step vertices for the construction of $G_f(i)$ from $f$.  
    Also, let the clusters of a final level step vertex of $G_f(i)$ be the two clusters added by the construction step that adds this step vertex.  
    Using our assumptions, it is easy to see that there exists an integer $i$ such that $G_f(i)$ has some two final level step vertices $v$ and $v'$ whose clusters, $T_1$ and $T_2$ for $v$ and $T_3$ and $T_4$ for $v'$, do not form a four-cycle.  
    Let $i$ be the largest integer with this property, and note that $G \neq G_f(i)$.  
    It is easy to see that neither $v$ nor $v'$ is a final level step vertex of $G_f(i+1)$.  
    Hence, wlog the $(i+1)^{st}$ construction step is $v'' \triangleleft (u \in T_1, w \in T_3)$.  

    Next, observe that the only clusters of $G_f(i)$ that contain $u$ and $w$ are the clusters of $v$ and $v'$, which do not form a four-cycle by assumption.  
    Hence, if no cluster of $v$ and cluster of $v'$, one containing $u$ and the other containing $w$, share a vertex, then Theorem \ref{the:four-cycle} shows that $G$ is not LCC.  
    Otherwise, wlog assume that $T_1$ and $T_3$ share a vertex $x$.  
    By the definitions of these clusters, any cluster of $G_f(i)$ not in $\{T_1,T_2,T_3,T_4\}$ either shares $x$ or no vertex with $T_1 \cup T_3$.  
    Therefore, no four-cycle of clusters in $G_f(i)$ contains $T_1$ and $T_3$, and so the same theorem shows that $G$ is not LCC.  
\end{proof}

For the next few results, consider the construction of a 1-dof tree-decomposable graph $G$ from any base nonedge $f$.  
Note that the clusters in the first two construction steps form a four-cycle.  
The diagonal pairs of this four-cycle that are not $f$ are the \emph{first level base nonedges} of $f$.  
For any such base nonedge $f'$, observe that $G$ can be constructed from $f'$ such that the first two construction steps create a four-cycle containing $f$ and then the remaining construction is identical to the construction from $f$ with the first two steps for $f$ left out.  

\begin{lemma}[1-Path LCC base nonedge exchange]
    \label{lem:1-path-lcc-base}
    Let $G$ be a 1-dof tree-decomposable graph that has at least six clusters and a 1-path construction from some base nonedge $f$.  
    If $G$ is LCC, then at least one endpoint of $f$ is contained in the final level for some first level base nonedge of $f$.  
\end{lemma}

\begin{proof}
    Assume to the contrary that $G$ is LCC but neither endpoint of $f$ is contained in the final level for any first level base nonedge $f'$ of $f$.  
    Consider the constructions of $G$ from $f$ and from $f'$, as discussed above.  
    First, we show that $G$ has a 1-path construction from $f'$.  
    Let $w_1$ and $w_2$ be the first two step vertices for $f'$.  
    Since $G$ has at least six clusters and a 1-path construction from $f$, neither endpoint of $f'$ and neither $w_1$ nor $w_2$ is a final level step vertex for $f$.  
    Hence, some final level vertex for $f$ is contained in $G \setminus \{w_1,w_2\}$.  
    Since neither endpoint of $f$ is contained in the final level for $f'$, neither is $w_1$ or $w_2$.  
    Also, since the constructions of $G$ from $f$ and from $f'$ are identical after their respective first two construction steps, a vertex of $G \setminus \{w_1,w_2\}$ is a final level step vertex for $f$ if and only if it is a final level step vertex for $f'$.  
    Combining these facts with our assumption that $G$ has a 1-path construction from $f$ shows that $G$ has a 1-path construction from $f'$.  

    Next, observe that $f'' = (w_1,w_2)$ is either $f$ or a first level base nonedge of both $f$ and $f'$.  
    We will show that at least one endpoint of $f''$ is contained in the final level for $f'$.  
    Note that the endpoints of $f'$ are first level step vertices for $f''$, and similarly the endpoints of $f''$ are first level step vertices for $f'$.  
    By Lemma \ref{lem:1-path-2-level-1}, these are the only first level step vertices for these base nonedges.  
    Recall the definitions of a final level step vertex of $G_f(i)$ and its clusters from the proof of this lemma.  
    Since $G$ has 1-path constructions from both $f'$ and $f''$, it is easy to see that either $G_{f'}(3)$ or $G_{f''}(3)$ contains some two final level step vertices whose clusters do not form a four-cycle.  
    Using an argument identical to that in the proof of Lemma \ref{lem:1-path-2-level-1}, we get that $G$ is not LCC, which contradicts our assumption.  
    Therefore, at least one endpoint of $f''$ is contained in the final level for $f'$.  

    Finally, it remains to consider the case where $f''$ is a first level base nonedge of $f$.  
    We can assume that $G_{f'}(1)$ and $G_{f'}(2)$ contains exactly one and exactly two endpoints of $f$, respectively, or else we can swap the roles of $f'$ and $f''$.  
    Since the first two step vertices for $f'$ are the endpoints of $f''$, one of which is a final level step vertex for $f'$, this shows that some endpoint of $f$ is contained in the final level for $f'$.  
    This completes the proof.  
    
\end{proof}

\begin{lemma}[1-Path LCC recursive structure] 
    \label{the:chain} 
    Let $G$ be a 1-dof tree-decomposable graph with at least six clusters, and consider its constructions from any base nonedge $f = (v_0,v'_0)$ and from any first level base nonedge $f'$ of $f$, as discussed above.  
    Also, let $X$ be the intersection of $\{v_0,v'_0\}$ and the final level vertex set for $f'$.  
    Lastly, let $G'$ be obtained from $G$ by performing all construction steps in the construction of $G$ from $f'$ except the ones that add at least one vertex in $X$.  
    If $G$ is LCC and has a 1-path construction from $f$, then $G'$ is LCC and has a 1-path construction from $f'$.  
\end{lemma}

\begin{proof}
    Assume $G$ is LCC and has a 1-path construction from $f$.  
    Since $f'$ is a base nonedge of both $G$ and $G'$, $G'$ is clearly LCC.  
    Let $w_1$ and $w_2$ be the first two step vertices for $f'$.  
    Since $G$ has at least six clusters and a 1-path construction from $f$, neither endpoint of $f'$ and neither $w_1$ nor $w_2$ is in the final level for $f$.  
    Also, since the constructions of $G$ from $f$ and from $f'$ are identical after their respective first two construction steps, a vertex of $G \setminus \{w_1,w_2\}$ is a final level step vertex for $f$ if and only if it is a final level step vertex for $f'$.  
    Furthermore, if some vertex in $\{w_1,w_2\}$ is in the final level for $f'$, then its construction step adds some endpoint in $X$.  
    These facts along with our assumption that $G$ has a 1-path construction from $f$ shows $G'$ has a 1-path construction from $f'$.  
\end{proof}

\begin{figure}[h]
	\centering
	\includegraphics[width=0.7\textwidth]{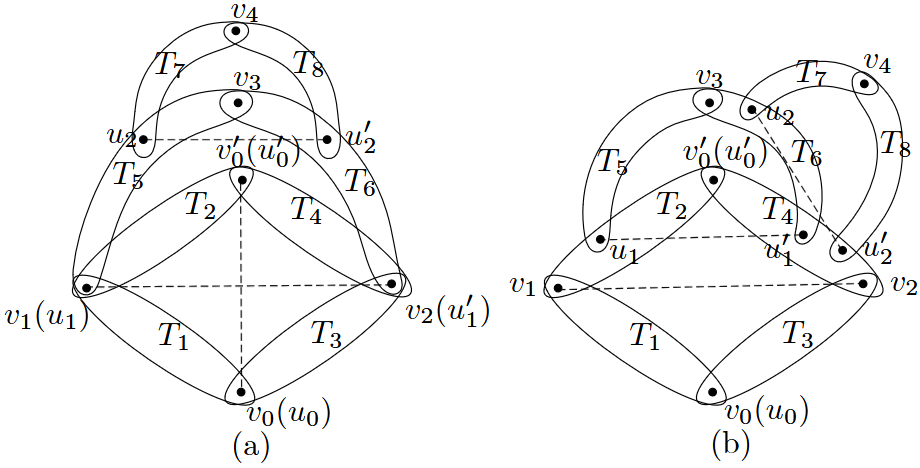} 
	
	\caption{Base pairs of vertices in Theorem \ref{thm:1-path-4-cycle}.  
    See also Section \ref{sec:Find-Cayley-configuration}.}
\label{F:quadrilaterals} 
\end{figure}

\begin{theorem}[1-Path four-cycle theorem]
    \label{thm:1-path-4-cycle}
    A 1-path graph with at least six clusters is nontrivially LCC if and only if for any 1-path construction of the graph from a base nonedge and any two consecutive constructions steps with distinct base pairs of vertices, there exists a four-cycle of clusters such that either the base pairs are the diagonal pairs of the four-cycle or one base pair is a diagonal pair and the other base pair is a chordal pair of the four-cycle (see Figure \ref{F:quadrilaterals}).  
\end{theorem}

\begin{proof}
    The converse direction follows immediately from Theorem \ref{the:four-cycle}.  
    For the forward direction, let $G$ be non-trivially LCC and have at least six clusters and a 1-path construction from some base nonedge $(v_0,v'_0)$.  
    We proceed by induction on the number $n$ of construction steps in this construction.  
    Since $G$ has at least six clusters, we have $n \geq 3$.  
    The theorem is easily verified when $n = 3$.  
    Assume the theorem holds when $n \leq k$, for any $k \geq 3$, and we will prove it when $n = k+1$.  
    By Lemma \ref{lem:1-path-lcc-base}, at least one endpoint of $f$ is contained in the final level for some first level base nonedge $f'$.  
    Consider the constructions of $G$ from $f$ and from $f'$, as discussed above.  
    In particular, the clusters added by the first two construction steps for $f$ form a four-cycle.  
    If the base pairs of vertices for these steps are distinct, note that one is a diagonal pair and the other is a chordal pair of the four-cycle.  
    
    Next, since the constructions of $G$ from $f$ and from $f'$ are identical after their respective first two construction steps, it suffices to show that all base pairs of vertices for $f'$ from the third one on satisfy the condition in the theorem statement.  
    Let $G'$ be the LCC graph with a 1-path construction from $f'$ obtained by applying Lemma \ref{the:chain}.  
    The definition of $G'$ and the fact that some endpoint of $f$ is contained in the final level for $f'$ imply that the construction of $G'$ from $f'$ has strictly fewer than $k+1$ construction steps.  
    Furthermore, this construction is identical to the construction of $G$ from $f'$, except one or two of the first two construction steps are left out.  
    Thus, applying the inductive hypothesis completes the proof.  
\end{proof}

\section{Theorem \ref{the:TriangleFreeCase}: Finite forbidden minor characterization of LCC 1-path graphs} 
\label{sub:Forbidden-minor-characterization}

Here we prove Theorem \ref{the:TriangleFreeCase} using Kuratowski's well-known finite forbidden minor characterization of planar graphs \cite{bib:Kuratowski}, below.  

\begin{theorem}
\cite{bib:Kuratowski} 
A graph is planar if and only if it has no $K_5$ or $K_{3,3}$
minor. 
\end{theorem}

The proof also requires Lemmas \ref{lem:4lemmas} and \ref{lem:equivalence}, below, which are proved using Lemma \ref{fact:noWellOn12}, also below.  

\begin{lemma}[Minimally rigid subgraphs do not contain base nonedges]
\label{fact:noWellOn12} 
No minimally rigid subgraph of a 1-dof tree-decomposable
graph $G$ contains both endpoints of any base nonedge of $G$.  
\end{lemma}

\begin{proof}
    As noted in Section \ref{sub:definitions}, for any base nonedge $f$ of $G$, $G \cup f$ is minimally rigid.  
    If some minimally rigid subgraph of $G$ contains both endpoints of $f$, then $G \cup f$ cannot be minimally rigid.  
    Combining these facts completes the proof.  
\end{proof}

\begin{figure}[h]
	\psfrag{v1}{$v_{0}$} \psfrag{v2}{$v_{0}'$} \psfrag{u1}{$v_{1}$}
	\psfrag{u2}{$v_{2}$} \psfrag{u3}{$v_{3}$} \psfrag{v5}{$v_{3}$}
	\psfrag{vn}{$v_{n}$} \psfrag{(a)}{(a)} \psfrag{(b)}{(b)} 
	
	\begin{centering}
	\includegraphics[width=0.6\textwidth]{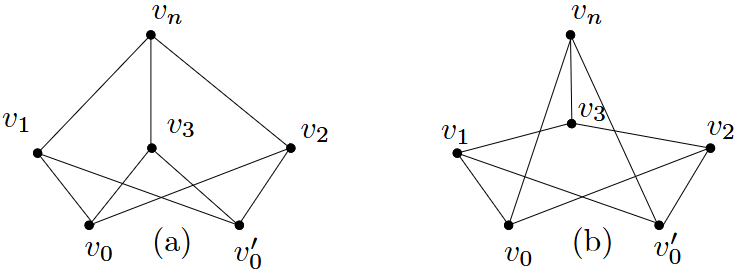} 
	\par\end{centering}
	
	\caption{Minors in the proof of Lemma~\ref{lem:4lemmas}.  
    }

    \label{F:4lemmas} 
\end{figure}

\begin{figure}[h]
	\psfrag{v0}{$v_{0}$}\psfrag{v0'}{$v_{0}'$}
	\psfrag{v1}{$v_{1}$}\psfrag{v2}{$v_{2}$}\psfrag{u}{$u$}\psfrag{w}{$w$}
	\psfrag{vk}{$v_{n}$}\psfrag{vc}{$v_{1}$} \psfrag{vn}{{$v_n$}}
	\psfrag{C1}{$C_{1}$}\psfrag{C2}{$C_{2}$}\psfrag{C3}{$C_{3}$}
	\psfrag{T1}{$T_{1}$}\psfrag{T1'}{$T_{1'}$}\psfrag{T2}{$T_{2}$}\psfrag{T2'}{$T_{2'}$}
	\psfrag{(a)}{(a)} \psfrag{(b)}{(b)} 	\psfrag{(c)}{(c)} 
	
	\begin{centering}
	\includegraphics[width=0.8\textwidth]{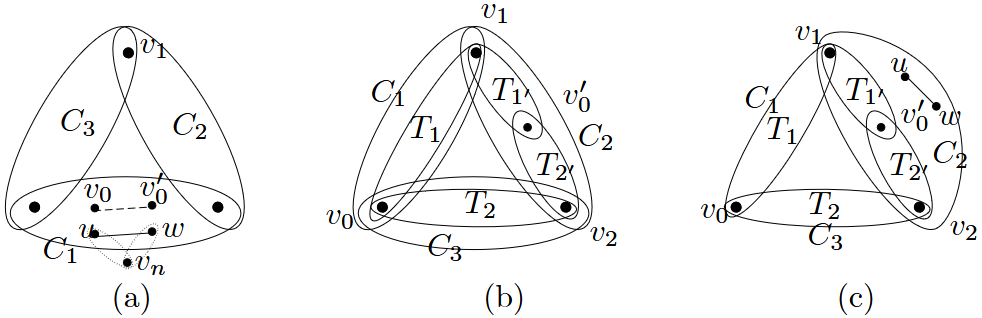} 
	\par\end{centering}
	
	\caption{Graphs for contradiction in the proof of Lemma~\ref{lem:4lemmas}.}

	\label{F:four-lemma-2} 
\end{figure}

\begin{lemma}[Bad $K_3$-free 1-path graphs] 
\label{lem:4lemmas} 
Let $G$ be a $K_3$-free 1-dof tree-decomposable graph that has a 1-path construction from a base nonedge $f$.  
If either $f$ has at least three first level step vertices or $f$ has exactly two first level step vertices and each endpoint of $f$ is contained in at least three clusters of $G$, then $G$ has a $K_{3,3}$ minor and is not LCC.  
\end{lemma} 

\begin{proof}
Let $f = (v_0,v'_0)$ and let $v_n$ be its only final level step vertex.  
First, we show that $G$ has a $K_{3,3}$ minor and is not LCC if $f$ has at least three first level step vertices $v_1$, $v_2$, and $v_3$.  
Clearly, $v_n$ is not in the first level for $f$.  
Consider the graph $G'$ obtained from $G$ by contracting all edges not incident to any vertex in $\{v_0, v'_0, v_1, v_2, v_3, v_n\}$ (see Figure~\ref{F:4lemmas}(a)). 
Since $G$ has a 1-path construction from $f$, $v_n$ must be adjacent to $v_1$, $v_2$, and $v_3$ in $G'$. 
Hence, $G'$ is clearly a $K_{3,3}$ minor of $G$ with $v_0$, $v'_0$ and $v_n$ in one partition and $v_1$, $v_2$ and $v_3$ in the other.

Next, assume that the extreme graph $\hat{G}_{f}(n)$ is tree-decomposable, and hence is the union of three tree-decomposable graphs $C_{1}$, $C_{2}$ and $C_{3}$ such that each pair shares exactly one vertex and all three shared vertices are distinct.  
If one of these graphs, wlog say $C_1$, contains both $v_0$ and $v'_0$, then Lemma \ref{fact:noWellOn12} shows that $C_1$ must also contain the extreme edge of $\hat{G}_{f}(n)$, as in Figure
\ref{F:four-lemma-2}(a).  
This implies that the vertex shared by $C_2$ and $C_3$ is a final level step vertex for $f$.  
However, this vertex is not $v_n$, which contradicts the fact that $G$ has a 1-path construction from $f$.  

Therefore, wlog assume that $C_1$ contains $v_0$ but not $v'_0$ and $C_2$ contains $v'_0$ but not $v_0$.  
Note that $C_3$ can contain at most one of these vertices, as in Figure \ref{F:four-lemma-2}(b).  
This figure also shows that the first level step vertices for $f$ are contained in $\{v_1,v_2\}$, which contradicts our assumption that $f$ has at least three such vertices.  
Consequently, $\hat{G}_{f}(n)$ is not tree-decomposable, and so $G$ is not LCC.  

Second, we show that $G$ has a $K_{3,3}$ minor and is not LCC if $f$ has exactly two first level step vertices $v_1$ and $v_2$ and each endpoint of $f$ is contained in at least three clusters of $G$.  
Recall from Section \ref{sub:definitions} that the clusters of a $K_3$-free 1-dof tree-decomposable graph are edges.  
Hence, since $G$ is $K_3$-free, the third construction step for $f$ must be $v_3 \triangleleft (v_1,v_2)$, as in Figure \ref{F:4lemmas}(b).  
Additionally, since $cdeg(v_0) \ge 3$, we have $v_3 \ne v_n$.  
Consider the graph $G'$ obtained from $G$ by contracting all edges not incident to any vertex in $\{v_0, v'_0, v_1, v_2, v_3, v_n\}$ (see Figure~\ref{F:4lemmas}(b)). 
Since $G$ is has a 1-path construction from $f$ and both $cdeg(v_0)$ and $cdeg(v'_0)$ are at least three, $v_n$ must be adjacent to $v_0$, $v'_0$, and $v_3$ in $G'$.  
Hence, $G'$ is clearly a $K_{3,3}$ minor of $G$ with $v_0$, $v'_0$ and $v_3$ in one partition and $v_1$, $v_2$ and $v_n$ in the other.  

Finally, assume that the extreme graph $\hat{G}_{f}(n)$ is tree-decomposable, and hence is the union of three tree-decomposable graphs $C_{1}$, $C_{2}$ and $C_{3}$ such that each pair shares exactly one vertex and all three shared vertices are distinct.  
As in the previous case, no graph $C_i$ can contain both $v_0$ and $v'_0$.  
Additionally, since $v_1$ and $v_2$ are in the first level for $f$, the shared vertices must be $v_1$, $v_2$, and one vertex in $\{v_0,v'_0\}$.  
Wlog, we can assume that $v_0$ is shared by $C_1$ and $C_3$ while $v'_0$ is contained in $C_2$, as in Figure \ref{F:four-lemma-2}(b).  
Observe that we can choose the first two construction steps for $f$ to be $v_1 \triangleleft (v_0 \in C_1, v'_0 \in T'_1)$ and $v_2 \triangleleft (v_0 \in C_3, v'_0 \in T'_2)$.  
Clearly, all clusters added by the remaining construction steps are contained in $C_2$, which does not contain $v_0$, and hence we get $cdeg(v_0) = 2$.  
However, this contradicts our assumption that $cdeg(v_0) \geq 3$.  
Thus, $\hat{G}_{f}(n)$ is not tree-decomposable, and so $G$ is not LCC.  
\end{proof}

\begin{lemma}[$K_3$-free 1-path recursive structure] 
\label{lem:equivalence} 
Let $G$ be a $K_3$-free 1-dof tree-decomposable graph that has at least four clusters and a 1-path construction from some base nonedge $f = (v_0,v'_0)$.  
Also, let $G'$ be $G \setminus \{v_0,v'_0\}$ if $v_0$ and $v'_0$ are each contained in at most two clusters, and let $G'$ be $G \setminus \{v_0\}$ if $v_0$ is contained in at most two clusters but $v'_0$ is not.  
In either case, $G'$ is $K_3$-free and has a 1-path construction from any first level base nonedge of $f$.  
\end{lemma}

\begin{proof}
    Since $G$ is $K_3$-free, neither does $G'$, and so we only need to show that $G'$ has a 1-path construction from any first level base nonedge $f'$ of $f$.  
    Note that $f'$ exists since $G$ has at least four clusters.  
    As discussed in Section \ref{sub:Characterizing-general}, $G$ has a construction from $f'$ that is identical to the construction the construction from $f'$ after the first two construction steps.  
    Recall from Section \ref{sub:definitions} that the clusters of a $K_3$-free 1-dof tree-decomposable graph are edges.  
    Hence, the endpoints of $f'$ are first level step vertices for $f$.  
    Since $G$ has a 1-path construction from $f$, it has exactly one final level step vertex, which is not an endpoint from $f'$.  
    These above facts show that a vertex of $G \setminus \{v_0,v'_0\}$ is a final level step vertex for $f$ if and only if it is a final level step vertex for $f'$.  
    Thus, $G'$ has a 1-path construction from $f'$.  
\end{proof}

We are now ready to prove Theorem \ref{the:TriangleFreeCase}.  

\begin{proof}[Proof of Theorem \ref{the:TriangleFreeCase}]
Let $G$ be a $K_3$-free 1-path graph and consider its construction from any base nonedge $f = (v_0,v'_0)$.  
Recall from Section \ref{sub:definitions} that the clusters of a $K_3$-free 1-dof tree-decomposable graph are edges.  
Let $m$ be the number of first level vertices for $f$, which are all step vertices by the previous fact.  
When $m \leq 1$, $G$ has exactly three vertices and is clearly is LCC and planar, and so the theorem holds.  
When either $m \geq 3$ or $m = 2$ and both $cdeg(v_0)$ and $cdeg(v'_0)$ are at least three, Lemma \ref{lem:4lemmas} shows that $G$ has a $K_{3,3}$ minor and is not LCC, and so the theorem holds.  
Hence, it remains to consider the case where $m = 2$ and wlog $cdeg(v_0) = 2$.  

Let $v_1$ and $v_2$ be the first level vertices for $f$.  
For the converse direction, we prove the contrapositive: if a $K_3$-free 1-path graph $G$ is not LCC, then it is not planar.  
In particular, we will show that $G$ has a $K_{3,3}$ minor.  
To the contrary, assume that $G$ has the minimum number of vertices such that it is LCC but does not have a $K_{3,3}$ minor.  
Let $G'$ be $G \setminus \{v_0,v'_0\}$ if $cdeg(v'_0) = 2$, and let $G'$ be $G \setminus \{v_0\}$ otherwise.  
Then, $G'$ is $K_3$-free and has no $K_{3,3}$ minor, since $G$ has neither of these.  
Note that $f' = (v_1,v_2)$ is a first level base nonedge of $f$, and so Lemma \ref{lem:equivalence} shows that $G'$ is $K_3$-free and has a 1-path construction from $f'$.  
Furthermore, $f'$ is a base nonedge of $G$, as discussed in Section \ref{sub:Characterizing-general}.  
Hence, since $G$ is LCC and its construction from $f'$ contains all the construction steps in the construction of $G'$ from $f'$, we get that $G'$ is LCC.  
Combining these facts shows that $G'$ contradicts our minimality assumption on $G$.  
Therefore, $G$ has a $K_{3,3}$ minor.  

\begin{figure}[h]
	\begin{centering}
	\psfrag{v3}{$v_{3}$} \psfrag{v1}{$v_0$} \psfrag{v2}{$v_0'$} 
	\psfrag{u1}{$v_{1}$} \psfrag{u2}{$v_{2}$} \psfrag{(a)}{(a)} \psfrag{(b)}{(b)} 
	\includegraphics[width=0.8 \textwidth]{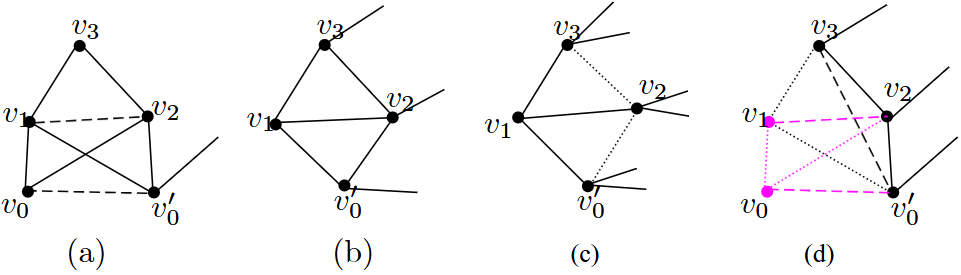} 
	\par\end{centering}
	
	\caption{Graphs in the proof of Theorem~\ref{the:TriangleFreeCase}.  
    The edge $(v_{0}, v_1)$ is contracted in (a) to get (b).  
    The edges $(v_2,v_3)$ and $(v_2,v_0')$ are deleted from (b) to get (c).
	The vertices $v_0$ and $v_1$ are deleted from (a) to get (d).}
	
	\label{F:theorem-contract} 
\end{figure}

Next, we prove the forward direction: if $G$ is LCC, then it is planar.  
Since each cluster of $G$ is an edge and $f$ has at least two first level step vertices, we have $cdeg(v'_0) \geq 2$.  
We prove the theorem for the case where $cdeg(v'_0) > 2$; the case where $cdeg(v'_0) = 2$ is similar so we omit the details.  
First, we show that $G$ has no $K_{3,3}$ minor.  
To the contrary, assume that $G$ has the minimum number of vertices such that it is LCC and has a $K_{3,3}$ minor.  
Note that the third base pair of vertices must be $(v_1,v_2)$, as in \ref{F:theorem-contract}(a).  
Also, as before, $G' = G \setminus \{v_0\}$ has strictly fewer vertices than $G$, is $K_3$-free, has a 1-path construction from $f'$, and is LCC.  
Hence, $G'$ cannot have a $K_{3,3}$ minor, or else $G'$ contradicts the minimality property of $G$.  
Consequently, the contrapositive of Lemma \ref{F:4lemmas} shows that some vertex in $\{v_1,v_2\}$, wlog say $v_1$, has degree 2 in $G'$, and hence it has degree 3 in $G$.  

Lastly, since $v_0$ has degree 2 and $K_{3,3}$ has no degree 2 vertices, either $(v_0,v_1)$ or $(v_0,v_2)$ must be contracted to reach any $K_{3,3}$ minor of $G$, as in Figure \ref{F:theorem-contract}(b).  
Additionally, $K_{3,3}$ is $K_3$-free, and so the two $K_3$ subgraphs in \ref{F:theorem-contract}(b) must be destroyed by some edge contractions and edge/vertex deletions.  
These operations must consist exactly of the two edge deletions resulting in the graph $H$ in \ref{F:theorem-contract}(c), or else we arrive at a similar contradiction.  
Let $G'' = G' \setminus\{v_1\}$ as in Figure \ref{F:theorem-contract}(d).  
Notice that $(v'_0,v_3)$ is a first level base nonedge of $f'$.  
Hence, since $v_1$ has degree 2 in $G'$, $G''$ is $K_3$-free and has a 1-path construction from $(v'_0,v_3)$.  
Consequently, the contrapositive of Lemma \ref{F:4lemmas} shows that some vertex in $\{v'_0,v_3\}$ has degree 2 in $G''$.  
Observe that $v_3$ has the same degree in $G''$ and $H$, and the same is true for $v'_0$. Therefore, one of these vertices has degree 2 in $H$.  
However, this implies that $H$ is a minor of $G'$, yielding a similar contradiction.  
Thus, $G$ does not have a $K_{3,3}$ minor.  

Finally, we show that $G$ does not have a $K_5$ minor.  
The argument is similar, up until we reach the graph $H$ in Figure \ref{F:theorem-contract}(b) by contracting $(v_0,v_1)$, in which $v_1$ has degree 3.  
since $K_5$ has no degree 2 vertex, the $K_5$ minor of $H$ must be reached by either deleting $v_1$ or contracting some edge incident to it.  
However, this implies that $G'$ has a $K_5$ minor, yielding a similar contradiction.  
Thus, $G$ does not have a $K_5$ minor.  
This completes the proof.  
\end{proof}

\begin{remark}
Planarity, LCC, and low Cayley algebraic complexity are all equivalent
for $K_3$-free 1-path graphs. 
\end{remark}


%

\subsection{Limits of finite forbidden minor characterization for LCC and low Cayley (algebraic) complexity}
\label{sec:tightness}

Here we prove Observation \ref{obs:GeneralTriCounter}.  

\begin{figure}[h]
	\centering
	\includegraphics[width=0.55\textwidth]{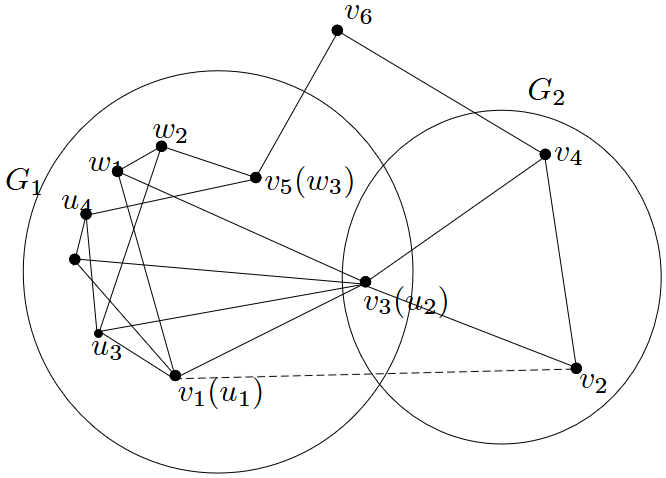} 
	
	\caption{A 1-path LCC graph with a $K_{5}$
	minor in the subgraph $G_{1}$.  
    See the proof of Observation~\ref{obs:GeneralTriCounter}.}

\label{F:TriangleFreeCounter1} 
\end{figure}

\begin{figure}[h]
	\centering
	\includegraphics[width=0.5\textwidth]{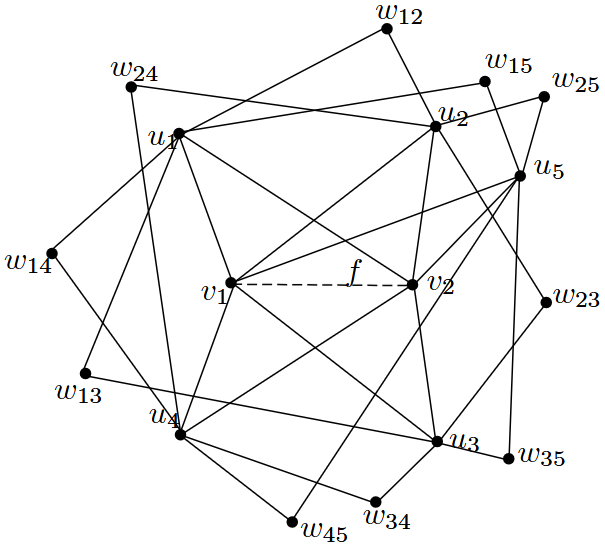} 
	
	\caption{A $K_3$-free LCC graph with a $K_{5}$ minor. 
    See the proof of Observation~\ref{obs:GeneralTriCounter}.  }

\label{F:OnePathCounter} 
\end{figure}

\begin{proof}[Proof of Observation \ref{obs:GeneralTriCounter}]
Let $G$ be a 1-path LCC graph, let $G'$ be a $K_3$-free LCC graph, and let $G''$ be a $K_3$-free 1-path graph.  
We will show that that we can choose $G$, $G'$, and $G''$ to additionally have a $K_n$ minor, for any $n \geq 1$.  
First, we prove this for $G$.  
Let $G$ have a 1-path construction from some base nonedge $f$.  
In particular, we show by induction on $n$ that we can choose $G$ to have a $K_n$ minor that contains the unique final level step vertex for $f$.  
The base case when $n = 1$ is trivial.  
Assume the claim holds when $n = k$, for any $k \geq 1$, and we will prove it when $n = k + 1$.  
Let $H$ be the 1-path LCC graph that has a $K_k$ minor given by the inductive hypothesis, and let it have a 1-path construction from some base nonedge $f'$.  
Also, let $u_1,\dots,u_k$ be vertices of $H$ that are mapped to distinct vertices of the $K_k$ minor, where $u_m$ is the unique final level step vertex for $f'$.  
We construct a graph $H'$ from $H$ by performing the following construction steps: $w_1 \triangleleft (u_1,u_2)$, $w_2 \triangleleft (w_1,u_3), \dots, w_{k-1} \triangleleft (w_{k-2},u_k)$.  
Let $G_1 = H' \cup f'$ and observe that it is tree-decomposable and contracting all its edges that are incident to any vertex other than $u_{1}$, $\dots$, $u_{k}$ and $w_{k-1}$ yields a $K_{k+1}$ minor.  
Let $G_2$ be a $K_3$ that contains $u_2$ and two vertices $x$ and $y$ not in $G_1$.  
Lastly, let $G$ be the 1-dof tree-decomposable graph obtained via the construction steps $u_2 \triangleleft (u_1 \in G_1, x \in G_2)$ and $v \triangleleft (w_{k-1} \in T_1, y \in T_2)$, for some tree-decomposable graphs $T_1$ and $T_2$.  
Then, $G$ clearly has a 1-path construction from $(u_1,x)$, where $v$ is the unique final level step vertex for $(u_1,x)$, is LCC, and has a $K_{k+1}$ minor that contains $v$.  
Refer to Figure~\ref{F:TriangleFreeCounter1} for a $K_{5}$ example.  
This proves the claim.

Next we prove the claim for $G'$.  
Let $G'$ be obtained from a nonedge $(v_1,v_2)$ by first performing the construction steps $u_1 \triangleleft (v_1,v_2),\dots,u_n \triangleleft (v_1,v_2)$ and then performing the construction steps $w_{ij}\triangleleft(u_{i},u_{j})$ for all $i\ne j$.  
Then, $G'$ is clearly $K_3$-free, and the LCC property can be verified using Theorem \ref{thm:1-path-4-cycle}.  
Furthermore, contracting all edges of $G'$ incident to any vertex other than $u_{1},\dots,u_{n}$ yields a $K_n$ minor (refer to Figure~\ref{F:OnePathCounter} for a $K_{5}$ example).  

Finally, clearly construction steps can be performed starting from $G'$ to obtain a graph $G''$ that is 1-path, $K_3$-free, and has a $K_n$ minor.  
Hence, by Theorem \ref{the:TriangleFreeCase}, $G''$ must not be LCC.  
This completes the proof.  
\end{proof}

\section{Theorem \ref{obs:k-path}: Algorithm for obtaining the Cayley configuration space for LCC linkage} 
\label{sec:Find-Cayley-configuration}

Here we prove Theorem \ref{obs:k-path}.  
First, we formally define minimal realization types, discussed in Section \ref{sec:contributions}.  
Consider the construction of an LCC graph $G$ from any base nonedge $f$.  
Note that every extreme edge of $G$ in this construction is a base nonedge of $G$.  

\begin{definition}[Reverse and minimal realization types]
Consider the construction of an LCC graph $G$ from any base nonedge $f$.  
A \emph{reverse realization type} for $f$ is a $T$-realization type containing the forward realization type for each extreme edge in this construction.  
A \emph{minimal realization type} for $f$ is a forward realization type for $f$ together with a reverse realization type for $f$.  
\end{definition}

For example,  for the linkage in Figure \ref{F:direction},  
realizations in (a) and (b) have different forward realization types (thus different minimal realization types). 
Moreover, since the underlying graph is LCC, 
the extreme graph $\hat{G}_{f}(3)$ has reverse construction $v_{0}\triangleleft(v_{1},v_{2})$, $v_{0}'\triangleleft(v_{1},v_{2})$,  
where (c)(e) and (d)(f) correspond to different reverse  realization types (thus different minimal realization types):
(c)(e) have $v_{0}$ and $v_{0}'$ on the same side of $(v_{1},v_{2}),$
while (d)(f) have them on opposite sides of $(v_{1},v_{2})$. 

Next, we present the QIM algorithm to compute a minimal oriented Cayley configuration space using Theorem \ref{thm:1-path-4-cycle} in Section \ref{sub:Characterizing-general}.



\begin{figure}[h]
	\centering
	{\tiny \includegraphics[width=\textwidth]{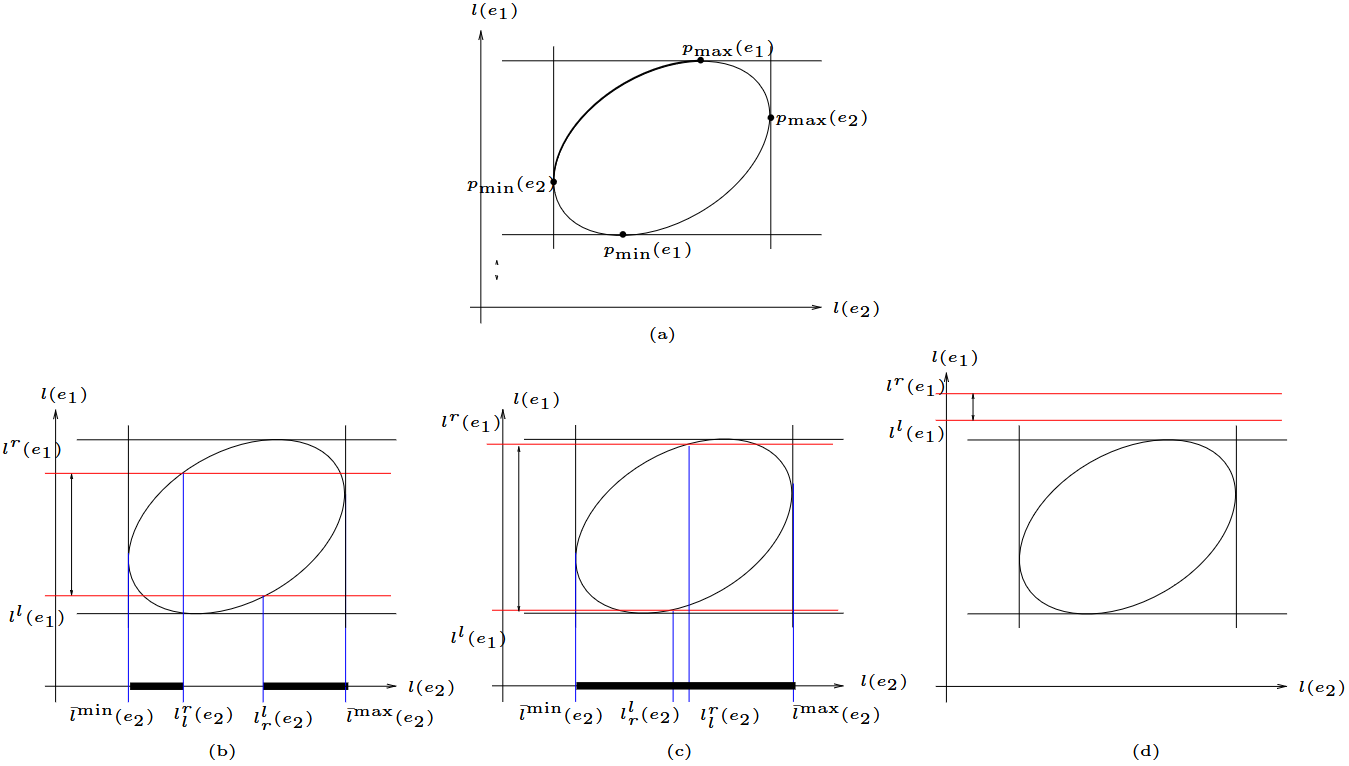} }
	
	\caption{(a) an example of ellipse $\mathcal{C}$ used in QIM, below; (b)(c)(d) various cases when mapping with ellipse $\mathcal{C}$.  }

\label{F:ellipse} 
\end{figure}

\medskip
\noindent\textbf{Algorithm (QIM)}: 
Let $(G, \bar{l})$ be a 1-path LCC linkage with a base nonedge $f$, let the construction steps from $f$ be ordered from $1$ to $k$, and let $f_i$ be the $i^{th}$ extreme edge for any $1 \leq i \leq k$.  
Also, let $I(f_i)$ be the interval $[\bar{l}^{\min}(f_i),\bar{l}^{\max}(f_i)]$ obtained by triangle inequality from extreme linkages of the extreme graph $\hat{G}_f(i)$.  
The algorithm computes the Cayley configuration space of $(G, \bar{l})$ over $f$, denoted by $S_1$.  
We start with the set $S_{k} \leftarrow \{I(f_k)\}$.  
If $k = 1$, we are done; otherwise, for each $i$ from $k-1$ to $1$, we show how to compute the set $S_i$.  

\smallskip
\noindent\textbf{Case 1:} $f_{i+1} = f_i$.  
\smallskip

Set $S_{i} \leftarrow S_{i+1} \cap I(f_i)$.

\smallskip
\noindent\textbf{Case 2:} $f_{i+1}$ and $f_i$ are diagonals of a four-cycle of clusters of $G$.  
\smallskip

Since we assume clusters have unique realizations, we can treat such the four-cycle as a quadrilateral with four sides $s_{1}$, $s_2$, $s_3$, $s_{4}$ and two diagonals $e_1$, $e_2$. 
For any 2D realization of this quadrilateral, the volume of the tetrahedron formed by $\{s_{1}$, $s_2$, $s_3$, $s_{4}$, $e_1$, $e_2\}$ must equal zero. 
Since we know the lengths $l(s_{1})$, $l(s_2)$, $l(s_3)$, $l(s_4)$, we can get from the volume equation (a so-called Cayley-Menger determinant) 
an implicit ellipse $\mathcal{C}$ relating attainable $l(e_1)$ and $l(e_2)$ values. 
See Figure \ref{F:ellipse} (a).  
So from the attainable interval $[l^{l}(e_1),l^{r}(e_1)]$ of one diagonal $e_1$, 
we can obtain the attainable intervals of $l(e_2)$ by mapping $[l^{l}(e_1),l^{r}(e_1)]$ on the curve $\mathcal{C}$, and vice versa.  
Figure \ref{F:ellipse} (b)(c)(d) illustrates several cases in determining the interval for $l(e_2)$ from $[l^{l}(e_1),l^{r}(e_1)]$.
Hence, setting $e_1 = f_{i+1}$ and its interval to $S_{i+1}$ and setting $e_2 = f_i$, we get $S_i$ via this map.  

\smallskip
\noindent\textbf{Case 3:} $f_{i+1}$ is a diagonal and $f_i$ is a chord of a four-cycle of clusters of $G$.  
\smallskip

Consider the case where $f_{i+1}$ and $f_i$ connect the same pair of adjacent clusters in the four-cycle.  
For example, in Figure \ref{F:quadrilaterals} (a), set $f_{i+1}=(u_2,u_2')$ and $f_i=(u_1,u_1')$.  
Consider the two triangles $\triangle v_3u_1u_1'$ and $\triangle v_3u_2u_2'$. 
Since $T_5$ and $T_6$ are fixed clusters, the lengths of triangle edges $(v_3,u_1)$, $(v_3,u_1')$, $(v_3,u_2)$ and $(v_3,u_2')$ are fixed. 
Moreover, if we know one of the two angles, $\angle u_1v_3u_1'$ and $\angle u_2v_3u_2'$, we can easily obtain the other. 
So from a value of $l(u_1,u_1')$, by the law of cosines, we can obtain $\angle u_1v_3u_1'$ and thus $\angle u_2v_3u_2'$, from which we can get a unique corresponding value of $l(u_2,u_2')$.  
Symmetrically, each value of $l(u_2,u_2')$ corresponds to a unique value of $l(u_1,u_1')$.  
$S_i$ is obtained via this map.  

Lastly, consider the case where $f_{i+1}$ and $f_i$ do not connect the same pair of adjacent clusters in the four-cycle.  
For example, in Figure \ref{F:quadrilaterals} (b), set $f_{i+1}=(u_2,u_2')$ and $f_i=(u_1,u_1')$.  
To obtain $S_i$, we can first map from $f_{i+1}$ to the diagonal $(v_3,u'_0)$ of the four-cycle, as above, and then to $f_i$ as in Case 2.  

\medskip

To prove Theorem \ref{obs:k-path}, we require Proposition \ref{obs:1-path}, below.  



\begin{figure}[h]
	\centering
	\includegraphics[width=0.6\textwidth]{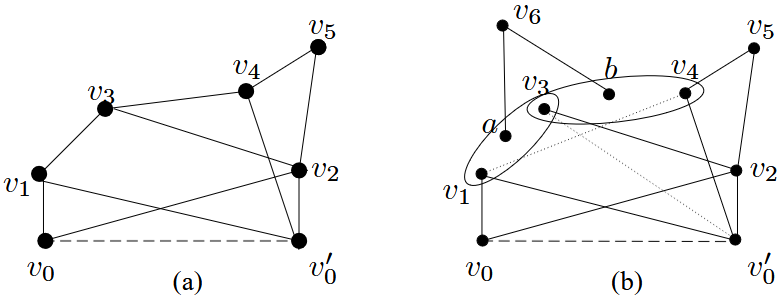} 
	
	\caption{(a) A 1-path graph with $v_5$ in the final level for $(v_0,v'_0)$.  
    (b) A tree-decomposable graph that is not 1-path.  
    See the discussion below.  }
\label{F:obstruction} 
\end{figure}

We give two examples to demonstrate how the QIM algorithm works. 

\smallskip
\noindent \textbf{Example 1: } 

To obtain the Cayley configuration space on $f=(v_{0},v_0')$ for $G$ in 
Figure~\ref{F:obstruction}(a):

\begin{itemize}
\item \textbf{Step 1:} 
Obtain the interval of $l(v_{4},v_2)$ in $\triangle v_2v_{4}v_{5}$ by triangle inequality; 

\item \textbf{Step 2:} In quadrilateral $v_0'v_2v_3v_4$, obtain
the interval of $l(v_0',v_{3})$ from the interval of $l(v_{4},v_2)$;

\item \textbf{Step 3:} Similarly, in quadrilateral $v_0'v_{1}v_{3}v_2$, 
we have $l(v_0',v_{3}) \rightarrow l(v_1,v_2)$;

\item \textbf{Step 4:} In quadrilateral $v_0v_2v_0'v_1$, 
we have $l(v_1,v_2) \rightarrow l(v_0,v_0')$. 
\end{itemize}

\begin{figure}[h]
	\centering
	\includegraphics[width=0.4\textwidth]{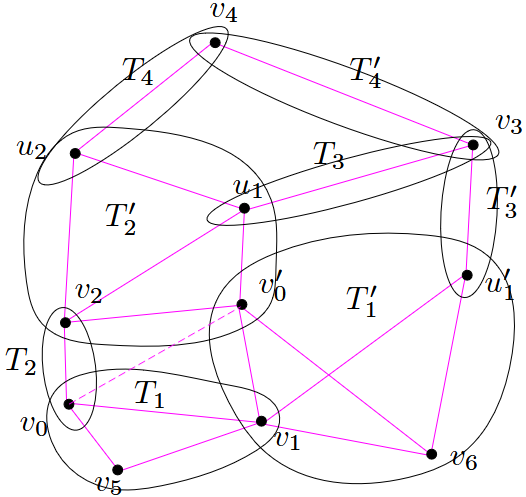} 
	
	\caption{Graph in Example 2, below.  }
\label{F:1PathOnly} 
\end{figure}

\noindent \textbf{Example 2: } 

To obtain the Cayley configuration space on $f=(v_{0},v_0')$ for $G$ in 
Figure~\ref{F:1PathOnly}: 

\begin{itemize}
\item \textbf{Step 1:} 
Obtain the interval of $l(u_2,v_3)$ in $\triangle v_4v_{3}u_2$ by triangle inequality; 

\item \textbf{Step 2:} In four-cycle $T_2'T_3T_3'T_1'$,
we have $l(u_2,v_3) \rightarrow l(u_1,u_1')$; 

\item \textbf{Step 3:}  In four-cycle $T_2'T_1'T_1T_2$,
we have $l(u_1,u_1') \rightarrow l(v_0,v_0')$.  	
\end{itemize}


\begin{proposition}[QIM for fixed minimal realization type and 1-path LCC]
\label{obs:1-path}
For a 1-path LCC linkage on $n$ vertices and with base nonedge $f$, the QIM algorithm can be used to obtain a minimal oriented Cayley configuration space over $f$, which is a single interval, in $O(n)$-time.  
\end{proposition}

\begin{proof}
Let $(G,\bar{l})$ be a 1-path LCC linkage with base nonedge $f$.  
First, we show that the set $S_1$ output by the QIM algorithm is the Cayley configuration space over $f$. 
We proceed by induction on the number $n$ of construction steps in the construction of $G$ from $f$.  
When $n = \leq 2$, the lemma is easily verified.  
Assume the lemma holds when $n \leq k$, for any $k \geq 3$, and we will prove it when $n = k + 1$.  
Let $f'$ be a first level base nonedge of $f$.  
Then, the graph $G'$ obtained by applying Lemmas \ref{lem:1-path-lcc-base} and \ref{the:chain} is LCC and has a 1-path construction from $f'$ that contains at most $k$ construction steps.  
Let $\bar{l}'$ be the restriction of $\bar{l}$ to $G'$.  
By the inductive hypothesis, the set output by the QIM algorithm for $(G',\bar{l}')$ and $f'$ is the Cayley configuration space over $f'$, which is the set $S_2$ in the running of this algorithm for $(G,\bar{l})$ and $f$.  
Using the triangle-inequality involved in Case 1 or the maps discussed in Cases 2 and 3 of the QIM algorithm to obtain $S_1$ from $S_2$, we see that a point is contained in $S_1$ if and only if it corresponds to some realization of $(G,\bar{l})$.  

Second, given a minimal realization type, the QIM algorithm can be modified so that the mapping process is restricted to lengths of extreme nonedges in realizations with this minimal realization type.  
With this modification, we will show that the set $S_i$ is a single interval in each step of the algorithm.  
The claim is immediate in the first step, i.e., $S_k$ is a single interval.  
We will examine each case of the algorithm and show that the map from $S_{i+1}$ to $S_i$ is bijective, and hence one set is a single interval exactly when the other set is.  
For Case 1, this is immediate.  

For Case 2, we make the following observations about the ellipse $\mathcal{C}$ used in the algorithm.  
First, one specific value of $l(e_1)$ can map to up to 2 distinct corresponding values of $l(e_2)$.
Second, the overall maximum and minimum points of $\mathcal{C}$, $p_{\min}(e_1)$, $p_{\min}(e_2)$, $p_{\max}(e_1)$ and $p_{\max}(e_2)$, each corresponds to a change in the local orientation of some triple of vertices containing three vertices among the endpoints of $e_1$ and $e_2$. 
For example, the upper
left segment of $\mathcal{C}$, from $p_{\min}(e_2)$ to $p_{\max}(e_1)$,
corresponds to the orientation where the vertices of $e_1$ lie on different sides of the line specified by $e_2$, 
and the vertices of $e_2$ lie on the same side of the line specified by $e_1$.  
Therefore, since the minimal realization type is fixed, the mapping in Case 2 is restricted to a monotonic segment of $\mathcal{C}$.  
Consequently, this map is bijective, as desired.  

Lastly, in the first subcase of Case 3, the claim from the discussion in this case.  
In the second subcase, the map is a composition of the maps in the first subcase and in Case 2.  
As just discussed, the first map is bijective.  
Using the locations of the extreme edges $f_{i+1}$ and $f_i$ in the four-cycle of clusters, it is easy to see that the second map is restricted to a monotonic segment of $\mathcal{C}$, as above, and hence is bijective.  

Third, the algorithm applies one map in each construction step, and we only need to apply this map to the endpoints of $S_{i+1}$ to get $S_i$.  
Since this takes constant time, the overall time-complexity is clearly $O(n)$.  
\end{proof}

Finally, we prove Theorem \ref{obs:k-path}. 

\begin{proof}[Proof of Theorem \ref{obs:k-path}]  
Consider an LCC linkage $(G,\bar{l})$ on $n$ vertices and with a base nonedge $f$, and let $\sigma$ be a minimal realization type for $G$ from $f$.  
For any final level step vertex $v$ for $f$, consider a minimal subsequence of construction steps in the construction of $G$ from $f$ whose last construction step adds $v$.  
Observe that the graph $G_v$ resulting from these construction steps is LCC and has a 1-path construction from $f$.  
Let $\bar{l}_v$ be the restriction of $\bar{l}$ to $G_v$ and let $\sigma_v$ be the restriction of $\sigma$ to triples in $G_v$.  
Consider the $\sigma_v$-oriented Cayley configuration space $I_v$ of $(G_v,\bar{l}_v)$ over $f$, which is a single interval and can be obtained via the QIM algorithm in $O(n)$-time, by Proposition \ref{obs:1-path}.  
We claim that the $\sigma$-oriented Cayley configuration space $I$ of $(G,\bar{l})$ over $f$ is the intersection of all intervals $I_v$.  
For any point $x$ not in this intersection, some final level step vertex $v$ for $f$ is such that $(G_v,\bar{l}_v)$ has no $\sigma_v$-oriented realization in which $f$ attains the length $x$.  
This implies that $x$ is not in $I$.  
Conversely, for any point $x$ in this intersection, each final level step vertex $v$ for $f$ is such that $(G_v,\bar{l}_v)$ has some $\sigma_v$-oriented realization in which $f$ attains the length $x$.  
Since $\sigma$ contains $\sigma_v$ and $\sigma_u$, for any two final level step vertices $v$ and $u$ for $f$, all of these realizations are consistent with each other in the sense that they can be combined to obtain a $\sigma
$-oriented realization of $(G,\bar{l})$.  
Therefore, $x$ is contained in $I$.  
Finally, there are at most $n$ intervals $I_v$, and so the overall time-complexity is $O(n^2)$.  
\end{proof}

\begin{remark} This result shows the advantage of the QIM algorithm over the ELR algorithm we mentioned in Section \ref{sec:Combinatorial-interpretation-of}. 
First, realizing each extreme linkage takes $O(|V|)$ time, 
therefore realizing all $O(|V|)$ extreme linkages from $f$ takes $O(|V|^2)$ time. 
Moreover, note that when the reverse realization type is fixed, 
an interval endpoint can also arise from a change in reverse realization type. 
This is not an extreme linkage for the given base nonedge, but an extreme linkage for an extreme nonedge. 
So the ELR algorithm must consider extreme linkages for all these $O(|V|)$ possible base nonedges, 
 and the overall time complexity is $O(|V|^{3})$.
\end{remark}

Why does the QIM algorithm fail as is,  
if the minimal realization type is not fixed? 
Notice that when the minimal realization type is not fixed, 
we cannot simply take the intersection of intervals as in proof of Theorem \ref{obs:k-path}, 
 since realizations requiring different minimal realization types can generate the same length for $f$. 
For example, refer to Figure \ref{F:obstruction} (b). 
There are two graphs $G_{v_5}$ and $G_{v_6}$. 
For certain $\bar{l}$, 
$G_{v_6}$ may require $v_1$ and $v_4$ to be on the same side of $(v_3,v_0')$ such that $v_6$ can be realized. 
At the same time, 
$G_{v_5}$ may require $v_1$ and $v_4$ to be on different sides of $(v_3,v_0')$ such that $v_5$ can be realized.
These two different realization types can generate the same length for $f$. 
Hence, if we just take the intersection of intervals $I_{v_5}$ and $I_{v_6}$, 
even if the intersection is non-empty, the linkage may still not have a minimal oriented realization.  

Given a minimal realization type, an alternative strategy for the QIM algorithm is to start with an interval $I_1$ for $f$ given by the first extreme linkages for which $f$ is the extreme edge.  
Then, for the $i^{th}$ construction step, take the interval $J$ for the $i^{th}$ extreme edge $e$ given by its extreme linkages, and set $I_i$ to the subset of $I_{i-1}$ of all lengths for $f$ that correspond to a realization where $e$ attains a length in $J$.  
This discussion leads to the following conjecture: 

\begin{conjecture*}[QIM without fixed minimal realization type]
Even when the minimal realization type is not fixed, 
the QIM algorithm can be adapted to work for LCC linkages.  
When the minimal realization type is fixed, 
this adapted algorithm should run in linear time in the number of vertices of the graph. 
\end{conjecture*}

\section{Theorem \ref{theo:path}: Finding a path of continuous motion between two realizations of an LCC linkage}
\label{sec:cont-path}

Here we prove Theorem \ref{theo:path} using Lemma \ref{lem:reachable}, below.



\begin{lemma}[Change of forward realization type] 
\label{lem:reachable}
Consider an edge-length preserving continuous motion path between two distinct realizations of a generic LCC linkage.  
For any base nonedge, the forward realization type changes at some realization along this path if and only if this realization corresponds to an interval endpoint in some forward oriented Cayley configuration space.  
Furthermore, at most one bit of the forward realization type can change at any point along the path.  
\end{lemma}

\begin{proof}
For any base nonedge $f$, assume the forward realization type changes at some realization $p$ along this path.  
Using Theorem \ref{lem:algebraic} (3), we see that the only changes that can occur are that some bits become zero and some bits become non-zero.  
Since the linkage is generic, exactly one bit changes, wlog say to zero.  
Let $x$ be the length of $f$ in $p$ and let $X$ any forward oriented Cayley configuration space whose forward realization type agrees with $p$.  
Note that $X$ contains $x$.  
Therefore, since $x$ is clearly the length of $f$ in a realization of some extreme linkage, it is an interval endpoint of $X$, by Theorem \ref{thm:interval-endpoints}.  

Next, assume the length of $f$ in some realization $p$ along the path is an interval endpoint in some forward oriented Cayley configuration space.  
By Theorem \ref{thm:interval-endpoints}, this endpoint is the length of $f$ in a realization of some extreme linkage.  
Hence, it is easy to see that any realization in the path sufficiently close to $p$ does not have the same forward realization type as $p$.  
\end{proof}

\begin{proof}[Proof of Theorem \ref{theo:path}]
Let $p$ and $q$ be two realizations of a generic LCC linkage, let $X$ be the Cayley configuration space of this linkage over any base nonedge $f$, and let $X(r)$ be the length of $f$ in any realization $r$ of this linkage.  
Also, let $Y_1$ be any the forward oriented Cayley configuration space whose forward realization type agrees with $p$, and let $I_1$ be the interval in $Y_1$ that contains $X(p)$.  
Theorem \ref{lem:algebraic} (3) can be used to obtain an edge-length preserving continuous motion path between $p$ and any realization $p_1$ such that $X(p_1)$ is an endpoint of $I_1$.  
By Lemma \ref{lem:reachable}, exactly one bit in the forward realization type of $p_1$ is zero, and wlog this bit is $1$ in the forward realization type of any point in the interior of $I_1$.  
Furthermore, there is a unique forward oriented Cayley configuration space $Y_2$ is such that its forward realization type differs from that of $Y_1$ only in this bit, which is $-1$ for $Y_2$, and $X(p_1)$ is an endpoint of some interval $I_2$ in $Y_2$.  

We can repeat the above process of traveling through forward oriented intervals and changing one bit of the forward realization type at a time until either an interval containing $q$ is found or we reach the minimum or maximum value in $X$.  
Another path can be obtained in a similar manner by traveling in the opposite direction starting from $p$.  
The time-complexity to find either path is clearly linear in the number of endpoints it contains.  
%
%
If $p$ and $q$ have the same minimal realization type, then Theorem \ref{obs:k-path} shows that $X(p)$ and $X(q)$ are contained in the same interval of the minimal oriented Cayley configuration space of the linkage over $f$, and hence a path between then can be found in constant time.  
\end{proof}

Figure \ref{F:tracking} gives an example 
of using the algorithm described in the above proof to 
finding an edge-length preserving continuous motion path
from realization (B1) with forward realization type $\sigma$ to (B2) with forward realization type $\tau$.  
We start from the interval $I_\sigma$ containing (B1) 
and take one endpoint of $I_\sigma$, which corresponds to extreme linkage realization (A1).  
Taking the entry of $\sigma$ which is 0 in (A1), and reversing its sign in $\sigma$, 
we get the next realization type $\tau$. 
Now we go from (A1) to (A2), 
which is essentially the same realization but contained in the oriented Cayley configuration space for realization type $\tau$, 
and the immediately reachable interval $I_\tau$ with (A2) as an endpoint realization is uniquely determined. 
Since $I_\tau$ contains the target realization (B2), a continuous path is successfully found.



\begin{remark}
    Given two Cayley configurations for a generic LCC linkage, we can apply the algorithm in the proof of Theorem \ref{theo:path} to attempt to find a path between some pair of corresponding realizations.  
    This potentially requires checking all pairs of forward realization types, and hence could take time exponential in the size oof the graph.  
\end{remark}


\section{Theorem \ref{thm:parameterize}: Bijectivity of representation}
\label{sec:ambient}

Here we prove Theorem \ref{thm:parameterize}.  
First, we give an algorithm to obtain a minimal bijective Cayley vector.  
which is based on the following characterization of graphs that are minimally globally rigid \cite{hendrickson1992conditions,jackson2005connected}.  

\medskip
\noindent\textbf{Algorithm to find minimal bijective Cayley vector}:
    Let $G$ be a 1-dof tree-decomposable graph with base nonedge $f$ such that each of its clusters is minimally globally rigid and each cluster that shares exactly two vertices with the rest of the graph is a single edge.  
    We construct a minimal bijective Cayley vector $F$ of $G$ iteratively as follows.  
    Start with $F_0 = \{f\}$.  
    \begin{enumerate}
        \item Let $F_1$ is a minimal-sized set containing $F_0$ and pairs of distinct final level step vertices for $f$ such that no pair of distinct final level step vertices for $f$ are separated by the endpoints of $f$ in $G \cup F_1$.  

        \item Note that $G \cup F_0$ is tree-decomposable, and hence $2$-connected.  
        Consequently, if $G \cup F_1$ is not $3$-connected, then it has a separator $U$ of size two.  
        By our assumption on the clusters of $G$ and the definition of $F_1$, $U$ does not contain both endpoints of $f$ or any final level step vertex for $f$.  
        Therefore, $U$ separates some endpoint $u$ of $f$ and some final level step vertex $v$ for $f$.  
        Let $F_2 = F_1 \cup \{(u,v)\}$.  

        \item Repeat Step (2) for $G \cup F_2$, and so on, until a $3$-connected graph $G \cup F_m$ is obtained.  
        Set $F = F_m$.  
        %
    \end{enumerate}

\begin{remark}
    If $G$ is 1-path in the above definition, then observe that $F$ has size two.  
\end{remark}

The proof of Theorem \ref{thm:parameterize} requires Lemma \ref{lem:globally_rigid_multi}, below, which is proved using Lemma \ref{lem:globally_rigid_1path}, also below.  
Both lemmas rely on the following two well-known reesults in combinatorial rigidity.  

\begin{theorem}[\cite{hendrickson1992conditions,jackson2005connected}]
    \label{thm:global-rigidity}
    A graph on at least three vertices is minimally globally rigid if and only if it is $3$-connected and redundantly rigid, i.e., deleting any of its edges yields a rigid graph.  
\end{theorem}

\begin{theorem}[\cite{PG,bib:Laman70}]
    \label{thm:rigidity}
    A graph $G = (V,E)$ is minimally rigid if and only if $|E|=2|V|-3$ and, for each subgraph $(V',E')$ of $G$, $|E'| \leq 2|V'|-3$.  
\end{theorem}

\begin{lemma}[1-Path global rigidity] 
\label{lem:globally_rigid_1path}
Let $G$ be a 1-path graph such that each of its clusters is minimally globally rigid and each cluster that shares exactly two vertices with the rest of the graph is a single edge.  
For any of its minimal bijective Cayley vectors $F$ obtained via the above algorithm, $G \cup F$ is minimally globally rigid.  
\end{lemma}

\begin{proof}
By Theorem \ref{thm:global-rigidity}, it suffices to show $G \cup F$ is 3-connected and redundantly rigid, which we do now.  
Note that $F$ contains exactly two nonedges: a base nonedge $f$ of $G$ and a nonedge $f'$ added in Step 2 of the algorithm.  
The construction of $F$ implies $3$-connectivity.  
Assume to the contrary that $G \cup F$ is not redundantly rigid, i.e., the graph $G_1$ obtained from $G \cup F$ by deleting some edge $e$ is not rigid.  
Since $G \cup f$ is minimally rigid, we have $e \neq f'$.  
Also, since all clusters are globally rigid, $e$ is either $f$ or a cluster.  

Next, since $G_1 = (V_1,E_1)$ is not rigid but $|E_1|=2|V_1|-3$, Theorem \ref{thm:rigidity} implies the existence of a proper subgraph $G_2 = (V_2,E_2)$ of $G_1$ with $|E_2|>2|V_2|-3$. 
The same theorem shows that $G_2$ must contain $f'$ since $G \cup f$ is minimally rigid.  
Hence, we have $G_2 = G_3 \cup f'$, where $G_3$ is a minimally rigid proper subgraph of $G_2$ containing $f'$ as a nonedge.  
Assume $G_3$ is contained in some cluster $T$.  
Since $G_3$ contains $f'$, one of whose endpoints $v$ is a final level step vertex for $f$, $T$ must share exactly two vertices with the rest of the graph.  
However, this implies that $T$, and hence $G_3$, is a single edge, contradicting the fact that $f'$ is a nonedge of $G_3$.  

Finally, note that $G_3$ is a subgraph of $G \cup f$ with $e$ deleted.  
If $G_3$ is not contained in any cluster, then it must contain $f$, since the only rigid subgraphs of $G$ with $e$ deleted are clusters.  
If $e = f$, then the facts that $v$ is the only final level step vertex for $f$ and $G_3$ is minimally rigid and contains $f$ and $f'$ imply that $G_3 = G$.  
This contradicts the fact that $G_3$ is a proper subgraph of $G_2$.  
Otherwise, if $e \neq f$, then $G_3$ does not contain some endpoint $x$ of $e$, or else $G_3 \cup e$ is a subgraph $G \cup f$ with more edges than is allowed by Theorem \ref{thm:rigidity}.  
Also, it is easy to show that any minimally rigid subgraph of a tree-decomposable graph is tree-decomposable.  
Hence, $G_3$ is tree-decomposable and can be constructed from $f$.  
However, this shows that some construction of $G$ from $f$ adds $v$ via a construction step prior to adding $x$, which contradicts the fact that $v$ is the only final level step vertex for $f$.  
This completes the proof.  
\end{proof}

\begin{lemma}[Multi-path global rigidity] 
\label{lem:globally_rigid_multi}
Let $G$ be a 1-dof tree-decomposable graph such that each of its clusters is minimally globally rigid and each cluster that shares exactly two vertices with the rest of the graph is a single edge.  
For any of its minimal bijective Cayley vectors $F$, $G \cup F$ is minimally globally rigid.  
\end{lemma}

\begin{proof}
    If $G$ is 1-path, then the lemma follows from Lemma \ref{lem:globally_rigid_1path}.  
    Otherwise, by Theorem \ref{thm:global-rigidity}, it suffices to show $G \cup F$ is 3-connected and redundantly rigid, which we do now.  
    The construction of $F$ implies $3$-connectivity.  
    Assume to the contrary that $G \cup F$ is not redundantly rigid, i.e., the graph $G_1$ obtained from $G \cup F$ by deleting some edge $e$ is not rigid.  
    Let $f$ be the base nonedge of $f$ contained in $F$.  
    Since $G \cup f$ is minimally rigid, we have $e \notin F \setminus f$.  
    Also, since all clusters are globally rigid, $e$ is either $f$ or a cluster.  
    
    Next, if $e$ is $f$, then let $v$ be any final level step vertex for $f$.  
    Otherwise, since some endpoint $x$ of $e$ is not a final level step vertex for $f$, there exists a final level step vertex $v$ for $f$ such that, in the construction of $G$ from $f$, $x$ must be added via a construction step prior to $v$ being added.  
    Assume some nonedge $f'$ in $F$ contains $v$ and some endpoint of $f$.  
    Since $G$ is not 1-path, there exists a proper subgraph $G'$ of $G$ that has a 1-path construction from $f$ such that $v$ is the only final level step vertex for $f$.  
    By Lemma \ref{lem:globally_rigid_1path}, $G' \cup F'$, where $F' = \{f,f'\}$, is minimally globally rigid.  
    Hence, deleting $e$ from $G' \cup F'$ yields a rigid graph.  
    Additionally, starting from $G' \cup F'$ with $e$ deleted and performing the construction steps between $G'$ and $G$ preserves rigidity.  
    However, this contradicts the fact that $G_1$ is not rigid.  

    Finally, assume no nonedge in $F$ contains $v$ and some endpoint of $f$.  
    Then, some nonedge $f'$ in $F$ contains $v$ and some other final level step vertex $u$ for $v$.  
    Clearly, there exists a subgraph $G'$ of $G$ that has a construction from $f$ such that $v$ and $u$ are the only final level step vertices for $f$.  
    Let $F' = \{f,f'\}$.  
    An argument as in the proof of Lemma \ref{lem:globally_rigid_1path} can be used to show that $G' \cup F'$ with $e$ deleted is rigid.  
    Therefore, as in the previous case, starting from $G' \cup F'$ with $e$ deleted and performing the construction steps between $G'$ and $G$ preserves rigidity.  
    This leads to the same contradiction, and so the proof is complete.  
\end{proof}

%

\begin{proof}[Proof of Theorem \ref{thm:parameterize}]
The bijectivity part of the theorem follows from Lemma \ref{lem:globally_rigid_multi}.  
The theorem follows from Lemma \ref{lem:globally_rigid_multi}.  
\end{proof}

Given a generic LCC linkage, we can visualize its realization space using the Cayley configuration space over a minimal bijective Cayley vector \cite{bib:beast}.  
This enables us to define a canonical distance between two connected components of the realization space,
and the distance separating two realizations not connected by an edge-length preserving continuous motion path \cite{bib:beast}.


\bibliographystyle{plain}

\appendix

\section{Proof of Theorem \ref{lem:algebraic} (structure of Cayley configuration space)}
\label{sec:Proof-for-lemma}

In the following, we denote the Cayley configuration space of a 1-dof tree-decomposable linkage $(G, \bar{l})$ over $f$ by $\Phi_{f}(G,\bar{l})$, and the oriented Cayley configuration space with forward realization type $\sigma$ by $\Phi_f(G, \bar{l}, \sigma)$.

\begin{proof}[Proof of Theorem \ref{lem:algebraic}]
We first prove for a fixed forward realization type $\sigma$,  
that the theorem holds for the oriented Cayley configuration space, 
by induction on the number of construction steps from $f$.  
In the base case, $G$ has only one construction step $v_{1}\triangleleft(v_{0}\in T_{1},v_{0}'\in T_{2})$.
The distances $\bar{l}(v_{1},v_{0})$ and $\bar{l}(v_{1},v_{0}')$ are
fixed by clusters $T_1$ and $T_2$ respectively, 
and by triangle inequality, 
$\Phi_{f}(G,\bar{l},\sigma)$
is a single closed interval $[|\bar{l}(v_{1},v_{0})-\bar{l}(v_{1},v_{0}')|,\bar{l}(v_{1},v_{0})+\bar{l}(v_{1},v_{0}')]$. 
Clearly,  (1) and (2) hold. 
For (3),  without loss of generality, 
let $p(v_0)$ be the origin, $p(v_0')$ lie on the $x$-axis, 
and $p(v_{1})=(x_{1},y_{1})$ has positive $y$-coordinate. 
Let $R_{1}=\bar{l}(v_{1},v_{0})$,
$R_{2}=\bar{l}(v_{1},v_{0}')$ and $R_{3}=l(v_{0},v_{0}')=l_f$.
In $\triangle v_0v_0'v_1$, we have

\[x_{1}=\frac{R_{1}^{2}+R_{3}^{2}-R_{2}^{2}}{2R_{3}} \cdot \sigma_1\]
\[y_{1}=\frac{\sqrt{(R_{1}+R_{2}+R_{3})(R_{1}+R_{2}-R_{3})(R_{1}-R_{2}+R_{3})(-R_{1}+R_{2}+R_{3})}}{2R_{3}}\]
where $\sigma_1$ is the entry corresponding to the first construction step in the forward realization type $\sigma$.  

Since the linkage is generic, $R_{3}$, namely $l_f$, cannot be $0$, 
so both $x_{1}$ and $y_{1}$ are continuous functions of $l_f$. 
Moreover, since internal realizations of both $T_{1}$ and $T_{2}$ are uniquely specified, 
the coordinates of all other vertices in $T_{1}$ and $T_{2}$ are continuous functions
of coordinates of $v_{1}$, $v_{0}$ and $v_{0}'$, thus continuous
functions of $l_f$. 

\smallskip{}

As induction hypothesis, 
assume that the theorem holds for linkages whose underlying graph $G_f(k-1)$ has $k-1$ construction steps.
Consider a graph $G_f(k)$ with $k$ construction steps, 
obtained by adding one more construction step $v_{k}\triangleleft(u_{k}\in T_{1},w_{k}\in T_{2})$
to $G_{f}(k-1)$. 
For any linkage $(G_f(k), \bar{l})$, according to Statement (3) of the induction hypothesis, 
$l(u_k,w_k)$ is a continuous function of $l_f$, say $l(u_k,w_k)=g(l_f)$. 
By triangle inequality,  $l(u_{k},w_{k})$ is restricted to the interval $[\min,\max]$
where $\min=|\bar{l}(u_{k},v_{k})-\bar{l}(w_{k},v_{k})|$ and $\max=\bar{l}(u_{k},v_{k})+\bar{l}(w_{k},v_{k})$.
This restriction may create new candidate interval endpoints in $\Phi_{f}(G_{f}(k),\bar{l}, \sigma)$,
namely $g^{-1}(\min)$ and $g^{-1}(\max)$, as shown in Figure~\ref{F:RccStepN}. 
A candidate endpoint is actually a new interval endpoint, 
only if its corresponding extreme linkage realization 
$p(\hat{G}_{f}(k),\bar{l}^{\min},\sigma)$ (resp. $p(\hat{G}_{f}(k),\bar{l}^{\max},\sigma)$) does exist. 
So (1) and (2) also hold for $(G_{f}(k),\bar{l})$.

\begin{figure}[h]
	\centering
	\includegraphics[width=0.35\textwidth]{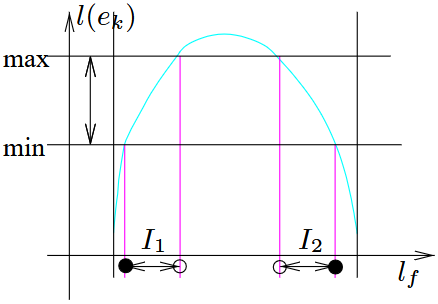} 
	
	\caption{See Theorem \ref{lem:algebraic}.  
    New constraint on $l(u_{k},w_{k})$
	creating interval endpoints in $\Phi_{f}(G_{f}(k),\bar{l}).$ $\bullet$:
	$g^{-1}(\min)$; { } $\circ$: $g^{-1}(\max)$. }

\label{F:RccStepN} 
\end{figure}

To prove (3), take any vertex $v$ in $G_{f}(k)$. 
By induction hypothesis, if $v\notin (T_{1}\cup T_{2})$, $p(v)$ is a continuous function of $l_f$. 
For $v\in (T_{1}\cup T_{2})$,  we first consider $p(v_{k})$. 
For convenience, first rotate and translate the coordinate system
so that $p(u_{k})$ is at the origin,  $p(w_{k})$ is on the $x$-axis, 
and $p(v_{k}) = (x_k, y_k)$ have positive $y$-coordinate. Let $R_{1}=\bar{l}(v_{k},u_{k})$,
$R_{2}=\bar{l}(v_{k},w_{k})$ and $R_{3}=l(u_{k},w_{k})$.
In $\triangle u_kw_kv_k$, we have

\[x_{k}=\frac{R_{1}^{2}+R_{3}^{2}-R_{2}^{2}}{2R_{3}} \cdot \sigma_k\]

\[y_{k}=\frac{\sqrt{(R_{1}+R_{2}+R_{3})(R_{1}+R_{2}-R_{3})(R_{1}-R_{2}+R_{3})(-R_{1}+R_{2}+R_{3})}}{2R_{3}}\]
where $\sigma_k$ is the entry corresponding to the $k^{th}$ construction step in the forward realization type $\sigma$.  

Since the linkage is generic, $R_{3}>0$. So both $x_{k}$ and $y_{k}$ are continuous functions of $l_f$. 
%
%
%
%
%
%
Moreover, since internal realizations for both $T_{1}$ and $T_{2}$
are specified, the coordinates of any other $v\in (T_{1}\cup T_{2})$ 
are continuous functions of $p(u_{k})$, $p(v_{k})$ and $p(w_{k})$, 
thus continuous functions of $l_f$. 
Consequently, for any nonedge $(u,w)$, 
$l(u,w)$ is also a continuous function of $l_f$.
Note that this continuity is not affected even if we transform back to the original coordinate system. 

For the complete Cayley configuration space, (1) and (2) still hold 
since $\Phi_f(G,\bar{l})$ is just the union of  oriented Cayley configuration spaces 
over all possible forward realization types.
\end{proof}

\section{Finding Cayley configuration spaces by realizing all extreme linkages (ELR)}
\label{sec:not-low}

Given a linkage $(G, \bar{l})$ and a forward realization type $\sigma$, 
in the ELR algorithm, 
we use a set $\mathcal{I}_{\sigma}$ to store candidate intervals of the oriented Cayley configuration space, 
which is initially the entire $\mathbb{R}^1$. 
For each construction step $i$, 
we update $\mathcal{I}_{\sigma}$ by considering restrictions on $l_f$ 
from all extreme linkages realizations of $\hat{G}_f(i)$ with forward realization type $\sigma$. 
After we have done this for every construction step, $\mathcal{I}_{\sigma}$ is the oriented Cayley configuration space. 

\medskip
\noindent \textbf{Algorithm (ELR):} 

\indent $\mathcal{I}_{\sigma}$ $\leftarrow$ $(-\infty, +\infty)$ \\
\indent \textbf{for} $i=1$ \textbf{to} $k$ \textbf{do}
		\indent [\textit{$k$ is the number of $G$'s  construction steps}] \\
	\indent \indent $S \gets \emptyset$
		\indent [\textit{set of candidate interval endpoints}] \\
	\indent \indent \textbf{for} every extreme linkage realization $p$ of $\hat{G}_f(i)$ with forward realization type $\sigma$\\
		\indent\indent\indent \textbf{if} $(G \cup f, \bar{l}\cup l_f)$ is realizable \\
			\indent\indent\indent\indent add $l_f$ value of $p$ to $S$ \\
	\indent \indent \textbf{for} each candidate endpoint $l_0$ in $S$ \\
	\indent \indent \indent UPDATE($\mathcal{I}_{\sigma}$, $l_0$) 
				\indent [\textit{see following discussion}] \\
\indent \textbf{return} $\mathcal{I}_{\sigma}$
\medskip

\begin{figure}[h]
	\centering
	\includegraphics[width=0.9\textwidth]{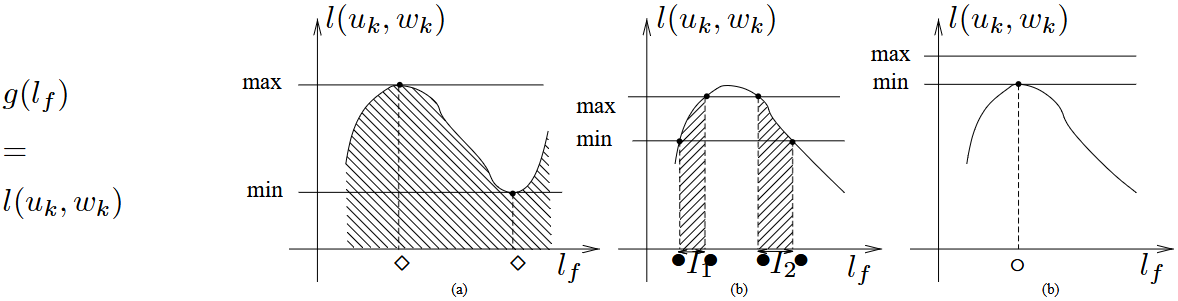} 
	
	\caption{See the discussion below.  
    $\min=|\bar{l}(v_{k},u_{k})-\bar{l}(v_{k},w_{k})|$, $\max=\bar{l}(v_{k},u_{k})+\bar{l}(v_{k},w_{k})$.
	(a) candidate $l_f$ values $\diamond$ that are internal points of some interval in $\mathcal{I}_{\sigma}$, not endpoints;
	(b) candidate $l_f$ values $\bullet$ that are new interval endpoints
	in $\mathcal{I}_{\sigma}$, creating intervals $I_{1}$ and $I_{2}$; 
	(c) candidate $l_f$ value $\circ$ that creates an isolated point in $\mathcal{I}_{\sigma}$. }

\label{F:endpoints} 
\end{figure}

When updating $\mathcal{I}_{\sigma}$, we need to notice that 
not every candidate Cayley configuration $l_0$ in $S$
actually creates new restriction on $\mathcal{I}_{\sigma}$.
Recall from the proof of Theorem \ref{lem:algebraic} that
a construction step $v_{i}\triangleleft(u_{i},w_{i})$
restricts $l(u_{i},w_{i})=g(l_f)$ in $[\min,\max]$. 
As shown in Figure~\ref{F:endpoints}, 
there are three possible cases 
for a candidate configuration $l_0$: 
(a) both the left and the right neighborhood of $l_0$ fall into $\mathcal{I}_{\sigma}$; 
(b) the left neighborhood of $l_0$ falls into $\mathcal{I}_{\sigma}$ but the right does not, and symmetrically,
the right neighborhood falls into $\mathcal{I}_{\sigma}$ but the left does not;
(c) neither the left nor the right neighborhood of $l_0$ falls into $\mathcal{I}_{\sigma}$, meaning that $l_f$ itself is the only realization in the neighborhood. 
In (b), $l_0$ creates an new endpoint in $\mathcal{I}_{\sigma}$. 
In (c), $l_0$ creates an isolated point in $\mathcal{I}_{\sigma}$. 
In (a), $l_0$ does not create any interval endpoint in $\mathcal{I}_{\sigma}$.

So in the UPDATE procedure, 

for each candidate Cayley configuration $l_0$, 
we check if there is any realization, with $l_f$ value between $l_0$
and the immediately preceding (resp. immediately succeeding) candidate interval endpoint 
in $\mathcal{I}_{\sigma}$.

\medskip
\noindent \textbf{Algorithm: UPDATE($\mathcal{I}_{\sigma}$, $l_f^{\min}$, $l_f^{\max}$)} 

\indent $prev \gets \max \{ l \in \mathcal{I}_{\sigma} | l < l_0 \}$, {   } 
		$next \gets \min \{ l \in \mathcal{I}_{\sigma} | l > l_0 \}$ \\ 
\indent $p \gets (prev+l_0)/2$, {   } $n \gets (next+l_0)/2$ \\ 
\indent $P \gets true$ if $p$ has corresponding realization, $false$ otherwise \\
\indent $N \gets true$ if $n$ has corresponding realization, $false$ otherwise \\
\indent \textbf{if} exactly one of $P$ and $N$ is $true$  \\
	\indent \indent add $l_0$ as an endpoint in $\mathcal{I}_{\sigma}$ \\
\indent \textbf{elseif} both $P$ and $N$ are $false$ \\
	\indent \indent add $l_0$ as an (isolated) endpoint to $\mathcal{I}_{\sigma}$ 
%
%

\noindent To obtain the complete Cayley configuration space, we just take $\Phi_{f}(G,\bar{l})=\bigcup\limits _{\sigma}\mathcal{I}_{\sigma}$.

\section{Observation \ref{obs: no-orientation}: Minimality of minimal realization type}
\label{sec:Exponential}

Here we prove Observation \ref{obs: no-orientation}.  



\begin{proof}[Proof of Observation \ref{obs: no-orientation}]
First, we give an example of an LCC linkage $(G, \bar{l})$ on $n$ vertices and
with a base nonedge $f$ and a fixed forward realization type $\sigma$ such that the $\sigma$-oriented Cayley configuration space $X$ of $(G,\bar{l})$ over $f$ contains exponentially many non-empty disjoint intervals in $n$. 
See Figure \ref{F:exponent-graph}. 
The base nonedge is $f = f_{1}=(v_{1},v_{3})$.
For convenience we slightly abuse our notation to let 
the construction step number start from $0$, 
so the $0^{th}$ and $1^{st}$ construction steps are 
$v_{4}\triangleleft(v_{1},v_{3})$
and $v_{2}\triangleleft(v_{1},v_{3})$ respectively. 
They form the outermost quadrilateral $Q_{1}=v_{4}v_{3}v_{2}v_{1}$.

For every $k>1$, 
the $k^{th}$ construction step  is $v_{k+3}\triangleleft(v_{k+2},v_{k})$,
which appends one vertex and two edges to the graph, and forms a nested
quadrilateral $Q_{k}=v_{k+3}v_{k+2}v_{k+1}v_{k}$. 
We denote the four
edges of $Q_{k}$ as $(v_{k+3},v_{k+2})=s_{k,1}$, $(v_{k+2},v_{k+1})=s_{k,2}$,
$(v_{k+1},v_{k})=s_{k,3}$ and $(v_{k},v_{k+3})=s_{k,4}$, 
and two diagonals $(v_{k+3},v_{k+1})=f_{k+1}$, $(v_{k+2},v_{k})=f_{k}$.
Notice that $Q_{k}$ shares two edges with $Q_{k-1}$: $s_{k,2}=s_{k-1,1}$, $s_{k,3}=s_{k-1,2}$.
The forward realization type $\sigma$ is assigned such that $v_{k+3}$ and $v_{k+1}$ lies on
different side of $f_{k}$. 

Clearly, $G$ is 1-path and LCC, and so we can use the QIM algorithm in Section
\ref{sec:Find-Cayley-configuration} to compute $X$. 
We start from the last extreme edge, 
and repetitively map $l(f_k)$ to get intervals for $l(f_{k-1})$, 
until we obtain the intervals for $l(f_1)$. 

\begin{figure}[h]
	\centering
	\includegraphics[width=0.35\textwidth]{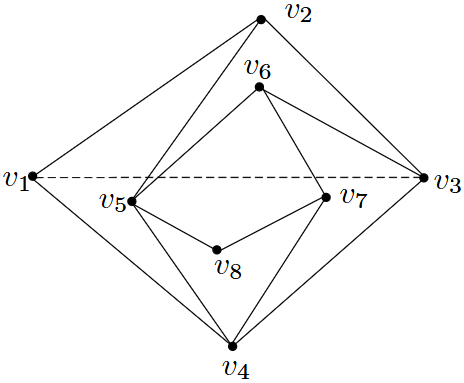}
	
	\caption{Graph in the proof of Observation \ref{obs: no-orientation}.  
	The graph is a series of nested four-cycles, and each cluster is
	an edge.}

\label{F:exponent-graph} 
\end{figure}

\medskip
\noindent \textbf{Example C.1}. 
We choose $\bar{l}$ such that for $Q_{1}$, $\bar{l}(s_{1,1})=\bar{l}(v_{4},v_{3})=8$,
$\bar{l}(s_{1,2})=\bar{l}(v_{3},v_{2})=8.1$, $\bar{l}(s_{1,3})=\bar{l}(v_{2},v_{1})=7.9$,
$\bar{l}(s_{1,4})=\bar{l}(v_{4},v_{1})=1$. 
The ellipse $\mathcal{C}$ relating
the two diagonals of $Q_{1}$ is shown in Figure \ref{F:diagonal-curve}. Only
the upper half of the curve (shown in solid line) corresponds to realizations
with forward realization type $\sigma$. 

For 
each $k>1$,
we assign $\bar{l}(s_{k,1})$ and $\bar{l}(s_{k,4})$
by the following: 
(1) 
Observe the ellipse $\mathcal{C}$ of $Q_{k-1}$, as shown in Figure \ref{F:diagonal-curve}. 
Denote the leftmost point on $\mathcal{C}$ by $p_{\min}(f_{k-1})$, the rightmost point by $p_{\max}(f_{k-1})$, the topmost point by $p_{\max}(f_{k})$. 
Let $l_2$ be the length of $f_k$ at $p_{\max}(f_{k})$, 
and $l_1$ be the larger one of the lengths of $f_k$ at $p_{\min}(f_{k-1})$ and $p_{\max}(f_{k-1})$. 
The interval $[l_1,l_2]$ is attainable by $l(f_k)$ in $G_f(k-1)$. 
(2) Assign $\bar{l}(s_{k,1})$ and $\bar{l}(s_{k,4})$ 
such that $\bar{l}(s_{k,1})-\bar{l}(s_{k,4})=\bar{l}^{\min}(f_k)=(1+\epsilon)l_1$,
$\bar{l}(s_{k,1})+\bar{l}(s_{k,4})=\bar{l}^{\max}(f_k)=(1-\epsilon)l_2$, 
where $\epsilon$ is a positive value small enough so that $\bar{l}(s_{k,1})$
and $\bar{l}(s_{k,4})$ have positive solutions. 
In this way, $l(f_k)$ is restricted to an interval $[\bar{l}^{\min}(f_k),\bar{l}^{\max}(f_k)]$
slightly tighter than $[l_1,l_2]$.

For example, we want to assign $\bar{l}$ for the $2^{nd}$ construction step $v_{5}\triangleleft(v_{4},v_{3})$.
As shown in Figure \ref{F:diagonal-curve}, in $Q_{1}$, $l_2=7.9+1=8.9$. 
For $p_{\min}(f_{k-1})$, $l(f_1) = 8-1 = 7$, the corresponding $l(f_2) \approx 8.36$. For $p_{\max}(f_{k-1})$, $l(f_1) = 8+1 = 9$, the corresponding $l(f_2) \approx 7.42$. 
Therefore $l_1=\max[8.36, 7.42] = 8.36$.
So let $\bar{l}^{\min}(f_2)=\bar{l}(v_{5},v_{4})-\bar{l}(v_{5},v_{3})=(1+10^{-5})l_1 \approx 8.364$,
$\bar{l}^{\max}(f_2)=\bar{l}(v_{5},v_{4})+\bar{l}(v_{5},v_{3})=(1-10^{-5})l_2 \approx 8.900$.
We assign $\bar{l}(v_{5},v_{4}) \approx 8.632$, $\bar{l}(v_{5},v_{3}) \approx 0.268$.
The two new extreme linkages $(\hat{G}_{f}(2),\bar{l}^{\min})$ and
$(\hat{G}_{f}(2),\bar{l}^{\max})$ each has two realizations, and each
of these realizations creates a new endpoint in  $\Phi_{f}(G,\bar{l})$:
$(\hat{G}_{f}(2),\bar{l}^{\max})$ corresponds to realizations in Figure
\ref{F:diagonal-curve} (b) and (d) , $(\hat{G}_{f}(2),\bar{l}^{\min})$
corresponds to realizations in Figure \ref{F:diagonal-curve} (a)
and (c) (point $b$, $d$, $a$ and $c$ in the left graph respectively).
Since $l_{b}(f_{1})\approx7.00$, $l_{d}(f_{1})\approx8.52$,
$l_{a}(f_{1})\approx7.49$, $l_{c}(f_{1})\approx7.51$,
$\Phi_{f}(G,\bar{l})$ contains two intervals $I_{1}=[7.00,7.49]$
and $I_{2}=[7.51,8.52]$, 
corresponding to two different reverse realization types. 


\begin{figure}[h]
	\centering
	\includegraphics[width=0.85\textwidth]{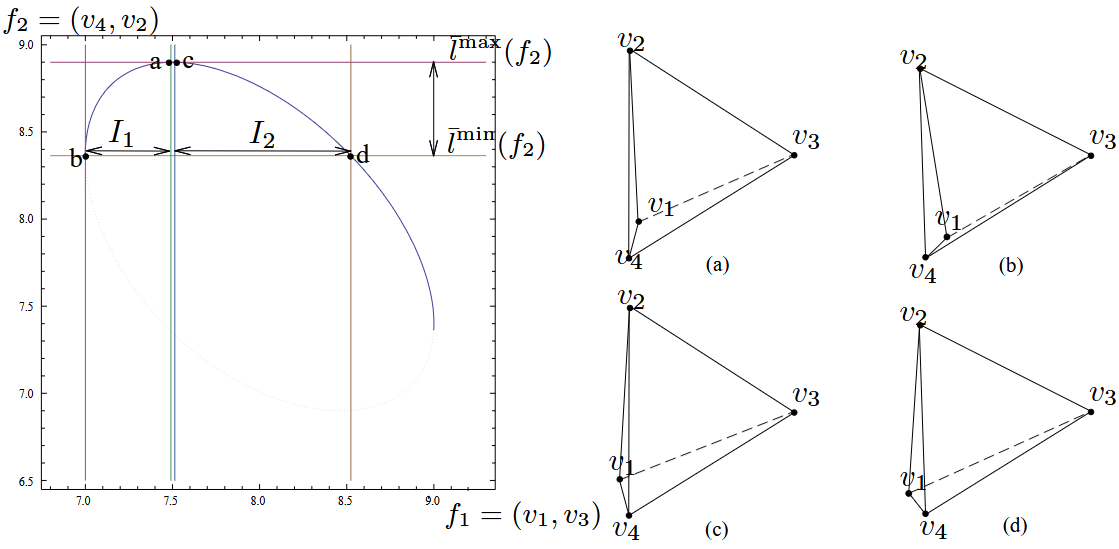} 
	
	\caption{See Example C.1.  
    Ellipse $\mathcal{C}$ for quadrilateral $Q_1 = v_{4}v_{3}v_{2}v_{1}$.
	Length of extreme edge $(v_{2},v_{4})$ is restricted by the 
	$2^{nd}$ construction step, and of $l(v_{1},v_{3})$ has
	2 intervals. }

\label{F:diagonal-curve} 
\end{figure}

Table \ref{T:edge-length} shows $\bar{l}$ for
the subsequent construction steps, computed by the procedure described above. 
Figure \ref{F:diagonal-curve-2} shows $\Phi_{f}(G_{f}(6),\bar{l})$. 
The single interval of $l(v_{5},v_{3})$ maps to 2 intervals
for $l(v_{2},v_{4})$: $I_{1}=[8.36,8.48]$, $I_{2}=[8.49,8.74]$,
and 4 intervals for the base nonedge $l(v_{1},v_{3})$:
$I_{11}=[7.000,7.008]$, $I_{21}=[7.010,7.121]$, $I_{22}=[8.039,8.391]$,
$I_{12}=[8.403,8.524]$. 

\begin{centering}
\begin{table}
	\begin{tabular}{|c|c|c|c|c|c|}
	\hline 
	$k$ & $\bar{l}(s_{k,1})$ & $\bar{l}(s_{k,2})$ & $\bar{l}(s_{k,3})$ & $\bar{l}(s_{k,4})$ & number of intervals for $l(v_{1},v_{3})$ after step $k$ \\
	\hline
	\hline 
	2 & 8.632 & 8 & 8.1 & 0.268 & 2 \\
	\hline 
	3 & 8.306 & 8.632 & 8 & 0.062 & 4 \\
	\hline 
	4 & 8.044 & 8.306 & 8.632 & 0.017 & 8\\
	\hline 
	5 & 8.645 & 8.044 & 8.306 & 0.004 & 16\\
	\hline 
	6 & 8.310 & 8.645 & 8.044 & 0.001 & 32\\
	\hline 
	7 & 8.045 & 8.310 & 8.645 & 0.0003 & 64\\
	\hline 
	8 & 8.645 & 8.045 & 8.310 & 0.00006 & 128\\
	\hline 
	9 & 8.310 & 8.645 & 8.045 & 0.00001 & 256\\
	\hline 
	10 & 8.045 & 8.310 & 8.645 & 0.000004 & 512\\
	\hline
	\end{tabular}
\medskip	
\caption{Example C.1. Edge lengths of quadrilateral $Q_{k}$ for construction step $2$ to $10$.}
\label{T:edge-length}
\end{table}
\end{centering}

\begin{figure}[h]
	\centering
	\includegraphics[width=0.7\textwidth]{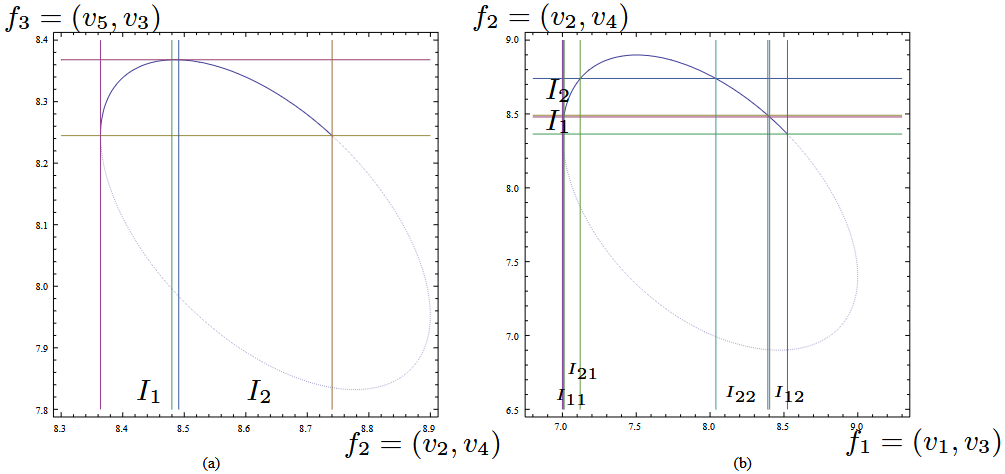} 
	
	\caption{See Example C.1.  
    Ellipse $\mathcal{C}$ for quadrilaterals (a) $Q_2 = v_{5}v_{4}v_{3}v_{2}$
	and (b) $Q_1 = v_{4}v_{3}v_{2}v_{1}$ after $v_{6}$ is realized. 
	The Cayley configuration space over $(v_1,v_3)$ is divided into 4 intervals $I_{11}$,
	$I_{21}$, $I_{22}$, $I_{12}$. }

\label{F:diagonal-curve-2} 
\end{figure}

In general, the $k^{th}$ construction step  $(k > 1)$ produces one interval for $l(f_k)$
which maps to 2 intervals for $l(f_{k-1})$, 4 intervals for $l(f_{k-2})$, and finally
$2^{k-1}$ intervals for $f_{1}=(v_{1},v_{3})$. 
Notice that there is no overlapping since 
the curve is monotonic in each interval.  

Finally, a symmetric argument shows that when $\sigma$ is a reverse realization type, $X$ can still have exponentially many non-empty disjoint intervals in $n$.  
\end{proof}

\end{document}